\documentclass[%
 aip,
 amsmath,amssymb,
preprint,%
]{revtex4-1}

\usepackage{graphicx}% Include figure files
\usepackage{dcolumn}% Align table columns on decimal point
\usepackage{bm}% bold math
\usepackage[mathlines]{lineno}% Enable numbering of text and display math
\usepackage[utf8]{inputenc}
\usepackage[T1]{fontenc}
\usepackage{mathptmx}
\usepackage{etoolbox}

\usepackage{xcolor}
\definecolor{customgreen}{HTML}{009933}
\newcommand*{\R}{\textcolor{red}}
\newcommand*{\B}{\textcolor{blue}}
\newcommand*{\G}{\textcolor{customgreen}}

\definecolor{darkpink}{rgb}{0.8, 0.2, 0.3}

\definecolor{olivegreen}{rgb}{0.33, 0.42, 0.18}

\begin{document}

%\preprint{Physics of Plasmas}

\title[Hybrid collisional-radiative modeling for high-fidelity atomic kinetics]{Hybrid collisional-radiative modeling for high-fidelity atomic kinetics}
% Force line breaks with \\
\author{Prashant Sharma}
\email{phyprashant@gmail.com}
\affiliation{Theoretical Division, Los Alamos National Laboratory, Los Alamos, NM 87545}

\author{Christopher J. Fontes}%
\affiliation{Computational Physics Division, Los Alamos National Laboratory, Los Alamos, NM 87545}

\author{Mark Zammit}
\affiliation{Theoretical Division, Los Alamos National Laboratory, Los Alamos, NM 87545}

\author{James Colgan}
\affiliation{Theoretical Division, Los Alamos National Laboratory, Los Alamos, NM 87545}

\author{Nathan Garland}
\affiliation{Theoretical Division, Los Alamos National Laboratory, Los Alamos, NM 87545}
\affiliation{Queensland Quantum and Advanced Technologies Research Institute, Griffith University, Nathan, Queensland 4111, Australia}
\affiliation{School of Environment and Science, Griffith University, Nathan, Queensland 4111, Australia}

\author{Xian-Zhu Tang}
\email{xtang@lanl.gov}
\affiliation{Theoretical Division, Los Alamos National Laboratory, Los Alamos, NM 87545}

%\date{\today}% It is always \today, today,
             %  but any date may be explicitly specified

\begin{abstract}
The fidelity of collisional-radiative (CR) models is critical for advancing our understanding of radiative properties and ionization balance in fusion plasmas. In this work, we present and evaluate hybrid CR schemes that combine fine-structure resolution with superconfiguration averaging, offering a practical compromise between accuracy and computational efficiency. Two hybrid CR models are developed for helium, lithium, and beryllium, retaining detailed fine-structure states up to selected principal quantum numbers, while higher-lying states are statistically averaged to form superconfigurations. These models are applied to compute radiative power loss, as well as average and effective charge states, across a wide range of electron temperatures and densities. The results are benchmarked against a fully fine-structure-resolved CR model to assess the accuracy of the hybrid approach. The findings demonstrate the versatility of hybrid CR schemes and their suitability for detailed plasma simulations where predictive fidelity must be balanced with computational cost.
\end{abstract}

\maketitle

\section{Introduction}

Atomic kinetic models are essential in the accurate modeling of
atomic-level populations and their evolution, which is crucial for
understanding a wide range of plasma phenomena. These models describe
the dynamical behavior of ionic charge states, excited states, and
electron distributions under diverse plasma conditions, and effects of
various atomic processes such as electron impact excitation and
de-excitation, ionization and recombination, radiative decay, etc. in
plasma evolution. By linking the microscopic details of atomic
collisions in terms of the atomic processes to the overall behavior of
the plasma, reliable atomic kinetic models can accurately predict the
radiative properties, ionization balance, and plasma transport
properties \cite{ralchenko2016modern}. Consequently, such models are
indispensable across a broad spectrum of applications, including
magnetic fusion \cite{reiter2005role,goto2014determination}, inertial-confinement fusion \cite{golovkin2006spectroscopic,scott2016collisional}, astrophysical \cite{pradhan2011atomic} and
atmospheric plasmas \cite{greenland2001collisional}, and laboratory low-temperature plasmas employed
in semiconductor processing \cite {donnelly2002optical} and plasma medicine applications \cite{akatsuka2019optical}.

Traditionally, plasma conditions have been approximated using simpler
atomic kinetic models. For high-density, thermally equilibrated
plasmas, local thermodynamic equilibrium (LTE) models are generally used, which assume that
atomic-level populations follow a Boltzmann distribution at the local
temperature, and that non-equilibrium processes can be ignored. At the
opposite extreme of very low densities and optically thin plasmas,
typical of astrophysical coronae, the coronal-equilibrium
(collisional-ionization) approximation is generally used, where the model assumes that spontaneous radiative
decay dominates the collisional processes. Although these
approximations greatly reduce computational demands, their accuracy is
limited to specific conditions, as they can fail in cases with
intermediate electron densities, complex ionizing or recombining
processes (e.g., charge exchange), or non-Maxwellian electron
distributions \cite{CHUNG20053}. Additionally, in most laboratory and
astrophysical plasmas, neither the coronal equilibrium limit nor the
LTE limit strictly applies. In these intermediate regimes,
collisional-radiative (CR) models provide a more general description
by simultaneously accounting for electron-impact processes, radiative
transitions, and finite-density effects. CR models bridge the
near-coronal and near-LTE regimes with fewer conceptual limitations
than those of simpler approaches. Indeed, in the low-density limit CR
models reduce to the coronal approximation, whereas at high densities,
they approach the LTE populations.

However, this generality and fidelity come at a significant
computational cost.  Detailed CR models can become computationally
demanding, especially when they include increasingly detailed atomic
structure data. For medium- and high-$Z$ elements, the number of
electronic configurations and fine-structure (FS) levels grows rapidly,
often scaling combinatorially with the principal quantum number $n$
and the total electron occupancy.  For example, going from a
configuration-average to a fully fine-structure-resolved model can increase the
number of coupled levels by one to two orders of magnitude.  The
resulting rate-equation matrix can easily exceed $10^{4}$ to $10^{5}$
levels for mid-$Z$ ions. Each additional
level adds new radiative decay channels, electron-impact excitation
and de-excitation processes, autoionization, and recombination rates
that must be tracked in a self-consistent manner.

As a result, the rate matrices become extremely large and dense,
straining memory and computation time, even when using advanced
solvers \cite{fontes2006large}. A fully fine-structure CR calculation
for a medium-$Z$ plasma may require gigabytes of memory and hours of
CPU time.  This rapid growth in complexity poses a major bottleneck
for detailed modeling of higher-$Z$ plasmas and motivates the search
for more computationally efficient approaches that preserve the
fidelity of key plasma observables, such as the average charge state,
radiative power loss, and charge-state distribution.

To address these computational challenges, we employ a hybrid CR scheme \cite{hansen2007hybrid} that combines fine-structure resolution with superconfiguration
averaging, applied to helium, lithium, and beryllium plasma
species. Two variants are developed for this study. The primary model,
Hybrid-($n_\text{valence}+4$), retains all states up to
$n_\text{valence}+4$ with full fine-structure resolution while
statistically averaging higher states (up to $n=10$), and a reduced
version, Hybrid-($n_\text{valence}$), is constructed to examine the
convergence behavior. Here, $n_\text{valence}$ denotes the principal
quantum number of the highest occupied shell in the ground state of
the atom or ion. A key distinction of our approach is that it begins
with a fully fine-structure level set and associated atomic data,
systematically grouping only the higher-lying states into
superconfigurations. This strategy significantly reduces the number of
levels and transitions, improving computational efficiency while
preserving essential atomic structure details. The hybrid framework
also provides a more general basis since collisional and radiative
rates are computed directly from fine-structure-resolved cross-sections and rates rather than from
pre-averaged quantities. Here we benchmark the hybrid model performance against
fully fine-structure-resolved CR models for the same species under steady-state
conditions, assessing the radiative power loss, average charge state,
and effective charge state across a wide range of electron
temperatures and electron densities.

\section{Collisional-radiative model}

The CR model describes the time evolution of atomic state populations
through a set of coupled rate equations that account for all processes
connecting different levels,
\begin{equation}
\frac{dN_i}{dt} \;=\; \sum_{j \neq i} N_j R_{j \to i} \;-\; N_i \sum_{j \neq i} R_{i \to j} \,,
\end{equation}
where $N_i$ and $N_j$ are the populations (or number densities) of atomic states
$i$ and $j$, respectively, and $R_{i \to j}$ denotes the rate for the
transition from state $i$ to state $j$ (with the inverse rate $R_{j \to i}$ defined
analogously). These rates include processes such as electron-impact excitation (EIE) and
de-excitation (EIDE), spontaneous radiative transitions, electron-impact ionization (EII), three-body recombination (3BR) and radiative recombination (RR), as well as other related processes. The accuracy of these rates depends critically on the
quality of the atomic data employed. For a reliable CR model, these
data should be benchmarked against high-quality experimental or
theoretical reference results to ensure that the relevant atomic
physics are accurately represented.

While the time-dependent formulation is crucial for understanding
transient plasma behavior, such as that arising from rapid changes in
plasma density, temperature, or incident radiation, many practical
situations allow for simplifications. Under conditions where the
plasma state changes slowly compared to the timescale of population
redistribution, a steady-state approximation is often employed. In
such regimes, the population distribution becomes time-independent, and
the rate equations simplify to
\begin{equation}
  \sum_{j \neq i} N_j R_{j \to i} \;-\; N_i \sum_{j \neq i} R_{i \to j} \;=\; 0.
\end{equation}
Solving the resulting algebraic system of equations allows for solution of atomic kinetics without the added
computational cost of the time integration. Such steady-state CR models
are routinely used for diagnostic purposes and spectral
interpretations, where long-lived or slowly varying conditions
prevail.

\subsection{Fusion Collisional-Radiative (FCR) Code}
In this work, we employ the FCR code, which has been recently
developed to investigate ionization balance and radiative properties
over a wide range of plasma conditions, with particular emphasis on
regimes relevant to fusion energy applications.  It supports multiple
refinement levels~\cite{fontes2016modern}, each designed to achieve an optimal balance between
computational cost and accuracy. For example, there is fine-structure
refinement for low-$Z$ elements, and hybrid or averaged
(configuration-average and superconfiguration) approaches for
intermediate- and high-$Z$ elements.

A key strength of FCR is its adaptive framework, which selectively
incorporates cross-sections and rate coefficients according to their
availability. The code preferentially uses cross-section data whenever
possible, enabling the treatment of arbitrary electron
energy-distribution functions. This feature makes FCR well suited for
both Maxwellian core plasmas and non-Maxwellian edge or boundary
plasmas, where phenomena such as momentum-space voids
\cite{guo-tang-prl-2012b,tang-guo-pop-2016,zhang-li-tang-sr-2024},
temperature anisotropy \cite{tang-guo-pop-2016}, bi-Maxwellian
distributions \cite{mao2023rapid}, and suprathermal electron
populations
\cite{mcdevitt-tang-guo-pop-2017b,mcdevitt-guo-tang-ppcf-2019a,garland2020impact}
may arise.

For rate calculations, FCR employs an adaptive Gauss–Kronrod quadrature method \cite{laurie1997calculation}, 
which extends the standard Gauss quadrature by adding extra integration points. This approach improves 
accuracy, particularly for functions with singularities or sharp gradients, without introducing 
significant computational cost. In the present study, all refinement schemes have been 
evaluated using a Maxwellian electron distribution function for clarity and consistency.

\subsection{Atomic data calculation and refinement schemes including hybrid scheme}

In the present work, the Flexible Atomic Code (FAC) \cite{FAC} is employed to generate a comprehensive set of atomic data, including level energies, oscillator strengths, spontaneous decay rates, and the relevant forward-process cross-sections and rate coefficients (EIE, EII, PI, and AI). These data are systematically incorporated into the FCR code to examine the radiative properties and ionization balance. It is important to note that the EIE cross-sections for neutral targets are calculated using the plane-wave Born (PWB) approximation, whereas the distorted-wave (DW) approximation is employed for ions. In contrast, other processes, including EII, PI, and AI, are treated using the DW method for all charge states, including neutrals.

This selection of methods provides a practical baseline for the present CR modeling and also establishes a flexible foundation for future refinements. In particular, the framework can readily incorporate cross-section scaling procedures for EIE \cite{kim1} and for EII \cite{jonauskas2020electron,jonauskas2022electron}, which allow neutral-target cross-sections to be refined to more closely reproduce results from non-perturbative approaches such as R-matrix \cite{bartschat1998r} and convergent close-coupling \cite{bray2017convergent}.

However, since the primary objective of this work is to establish and examine the hybrid CR modeling strategy, the discussion here remains focused on the refinement schemes themselves and their implications for atomic kinetics.

As discussed earlier, the CR models can be composed of states of
varying levels of refinement, reflecting a trade-off between accuracy
and computational feasibility. At the coarsest level, a
superconfiguration CR model \cite{CHUNG20053} groups energy states
solely by the principal quantum number ($n$), providing a compact
representation suitable for cases where only low-resolution data are
required, or computational resources are limited. A more refined
option is the configuration-average CR model
\cite{fontes2016modern,Fontes_2015,summers2005atomic,abdallah2009reduced}, which classifies states by both
$n$ and the azimuthal quantum number ($l$), thus capturing more
structural detail without the full complexity of resolving every
fine-structure level. The fine-structure ($J$-resolved) model
\cite{ralchenko2001accelerated,fontes2016modern,Fontes_2015,summers2005atomic} achieves 
the highest practical level of refinement by accounting for the coupling of
angular momenta associated with the orbital electrons, leading to a
total angular momentum $J$ for each atomic energy level.

\begin{figure}[!h]
\centering
\includegraphics[scale=0.25]{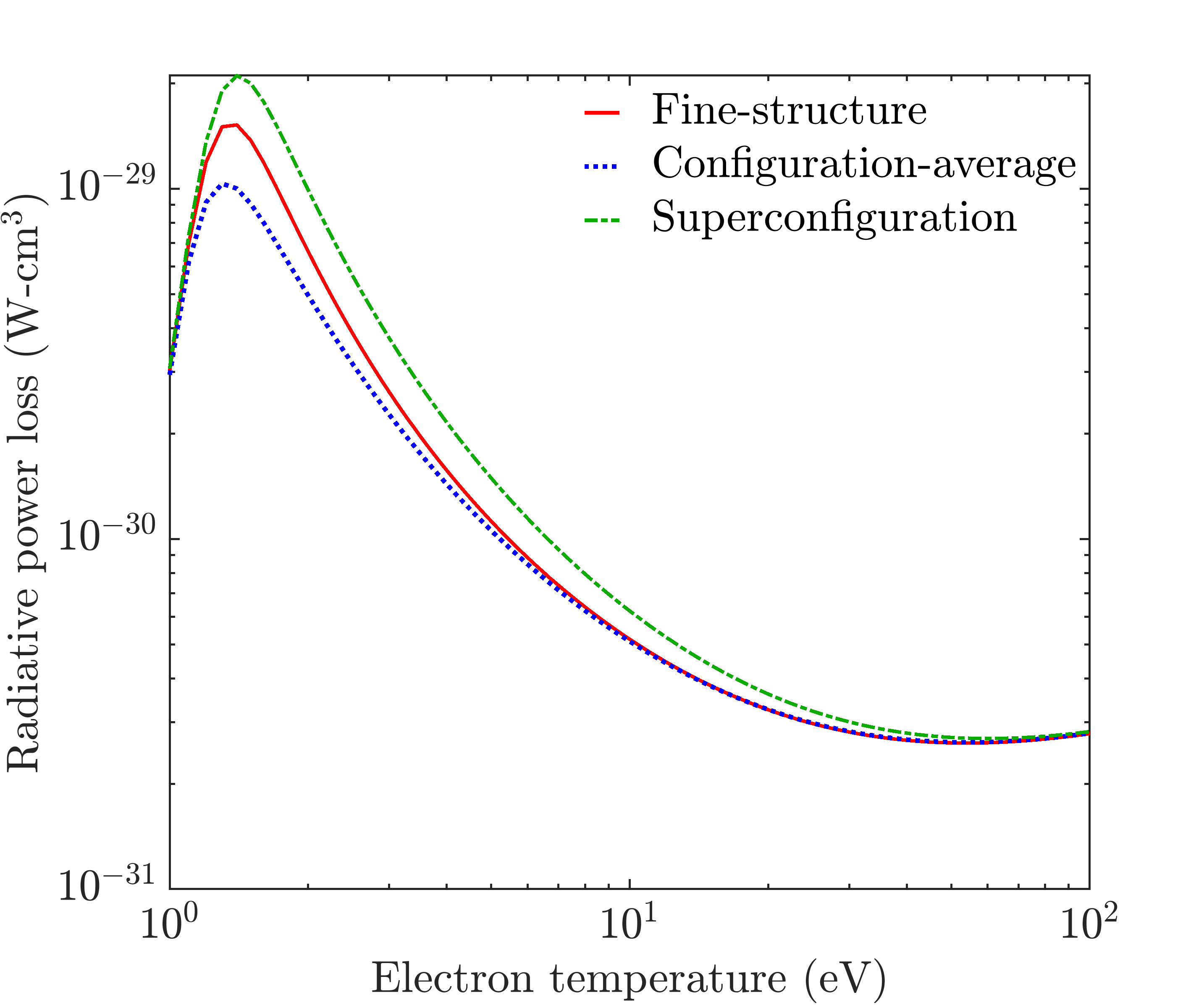}
\caption{\label{fine_config} \textbf{Hydrogen:} Comparison of
  radiative power loss calculated using the superconfiguration~(\G{$-\cdot$}),
  configuration-average~(\B{$\cdots$}), and fine-structure~(\R{--}) CR models for $n_e =
  10^{12}$ cm$^{-3}$. It is noteworthy that the superconfiguration and
  configuration-average CR models shown here are constructed through
  statistical averaging of the underlying fine-structure states and
  the atomic processes connecting them.}
\end{figure}

To highlight the impact of various refinement schemes on key plasma
properties, we present a comparison of the radiative power loss for
hydrogen at an electron density of $10^{12}$ cm$^{-3}$, obtained from
superconfiguration, configuration-average, and fine-structure CR
models, as shown in Fig.~\ref{fine_config}. In the present case of hydrogen, it is observed that the configuration-average model yields radiative power
losses that are approximately 50\% lower than those of the fine-structure model near the peak of the curve. This discrepancy originates from a “state-ordering problem”, i.e. the energy ordering of the fine-structure levels (for example, for $n=2$: $E_{2p_{1/2}} <  E_{2s_{1/2}} < E_{2p_{3/2}}$) cannot be preserved under configuration-averaging (which results, for example, in the following energy order for $n=2$: $E_{2s} < E_{2p}$). In addition (continuing with the $n=2$ example for definiteness), the energy location of the $2s_{1/2}$ level in between the $2p_{1/2}$ and $2p_{3/2}$ levels means that the $2s_{1/2}$ level can undergo both excitation and de-excitation processes with the two levels that arise from the $2p$ configuration. This situation introduces an ambiguity when trying to compute the configuration-average rates and cross sections via the configuration-average analog of the superconfiguration expression displayed in Eq.~(\ref{ss_cross}) below. In order to address this ambiguity, we choose to exclude the EIE cross-section and radiative decay rate for the $2p_{1/2} \leftrightarrow 2s_{1/2}$ transition when computing the configuration-average quantities.
With this choice, the $2s_{1/2}$ level, which has a dipole allowed transition with a lower level in the fine-structure model, becomes the $2s$ metastable, or forbidden, configuration in the configuration-average model, which results in a significant rearrangement of the populations between the two models at low densities. The $2s$ metastable configuration retains a significant amount of population that is not available for dipole emission, thus resulting in a decrease in the peak of the radiative power loss displayed in Fig.~\ref{fine_config}. In more complex multi-electron systems, the configuration-averaging can be more complicated than the hydrogenic case in the sense that both resonance and metastable levels can arise from the same configuration. In that case, configuration-average models combine the population of metastable levels with resonance levels that produce dipole radiation, which typically leads to higher, rather than lower, radiative power losses in configuration-average CR models~\cite{clark2003comparison}.

Analogous to the previous discussion for complex multi-electron systems, we note from Fig.~\ref{fine_config} that the superconfiguration model
shows roughly 30\% higher radiative power loss compared with the
fine-structure model. This overestimation is due to the grouping of all of the metastable, or forbidden, levels arising from a given Layzer complex~\cite{layzer1959} with the corresponding resonance levels in that complex. Since the metastable levels can possess a significant amount of population at low plasma densities, grouping them with the resonance levels can artificially increase the emission from dipole allowed transitions. In a fully fine-structure scheme, these metastable levels are treated explicitly and would contribute less to the
total radiative power loss. Consequently, while both
configuration-average and superconfiguration approaches offer
computational advantages over a fine-structure-resolved CR model, they
can introduce systematic errors in predicted plasma properties like
radiative power loss and average charge state. For example, in this
  specific case, discrepancies up to 30--50\% in the radiative power loss
  would have dominated the uncertainties in collisional plasma
  energy transport of a low-temperature boundary plasma.  This issue can be
  a particularly important concern in both: (i) designing plasma power exhaust
  and divertor detachment solutions for fusion power reactors, and (ii)
 quantitatively interpreting current experiments.  For practical
  applications like fusion energy, physics fidelity must be considered
  when choosing an appropriate level of refinement for the CR
  modeling.

To optimize the balance between completeness and computational tractability, we adopt a hybrid CR modeling strategy that shares the same underlying principle as Hansen et al. \cite{hansen2007hybrid}, though employing a distinctly different methodology. The approach by Hansen et al. begins
with a complete set of relativistic configuration-average (RC) states
and rates, then selectively replaces a subset of coronal
configurations with detailed FS levels, while averaging the remaining
RCs into non-relativistic configurations or superconfigurations. This
top-down construction ensures completeness, but relies on RC data as
the starting point. In contrast, the present work starts from a fully
fine-structure-resolved set of levels and systematically aggregates only
the highly excited or continuum-like (described below) states into
superconfigurations. This bottom-up construction retains full FS
detail for all low-lying states, which is crucial for describing
metastable populations, ladder ionization, and density-sensitive line
ratios, while adaptively reducing the state space until convergence of
key plasma quantities, such as charge-state distribution and radiative
power loss, is achieved.

\begin{figure}[!h]
\centering
\includegraphics[scale=0.12]{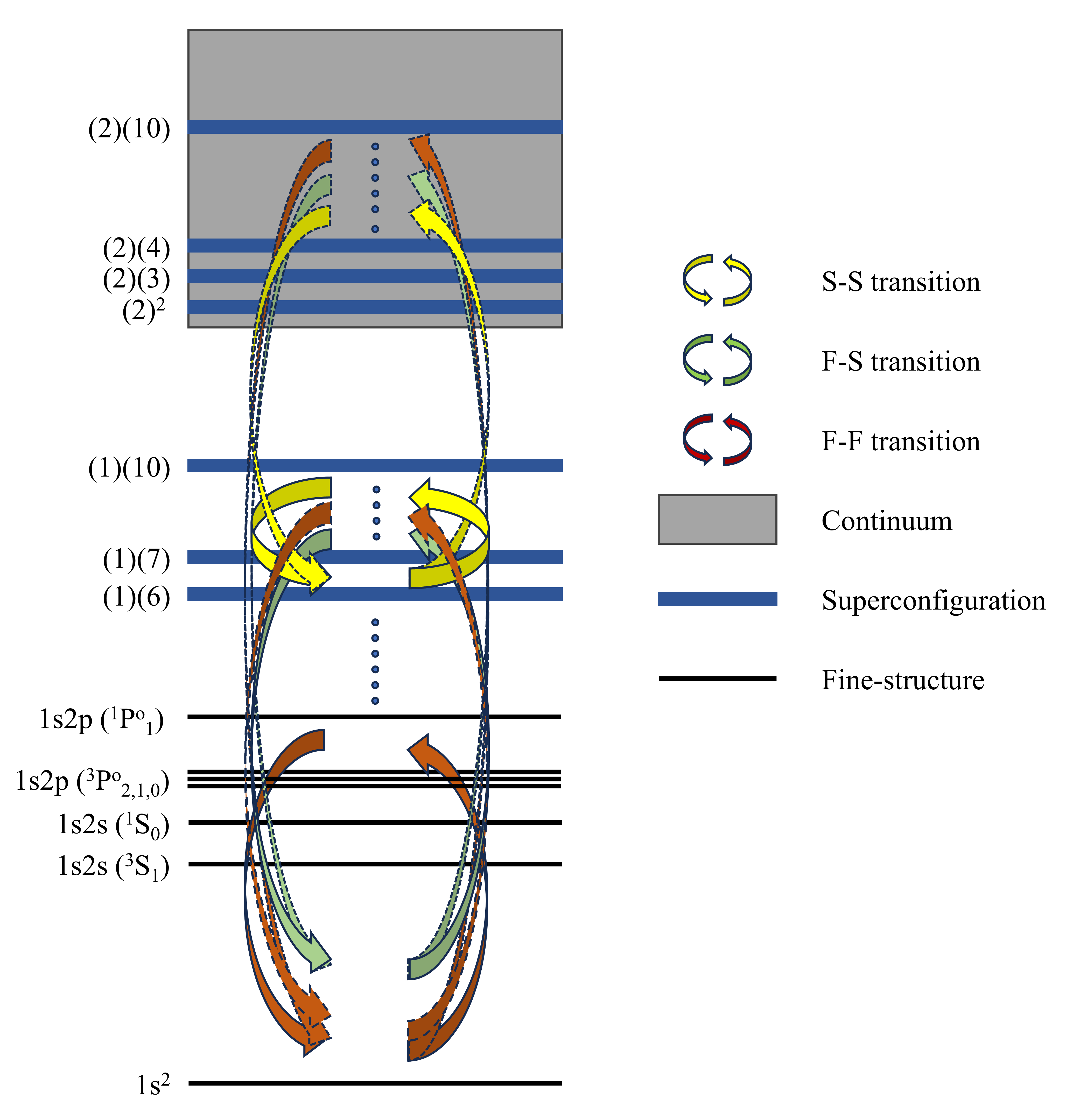}
\caption{\label{hybrid_schematic2} Schematic of the hybrid scheme for helium (or helium-like ions), showing transitions among fine-structure states, superconfiguration states, and the ``handshake" coupling that connects these two regimes. The lower-lying levels are treated with full fine-structure resolution, while higher-lying and continuum-like states are grouped into superconfigurations.}
\end{figure}

Figure~\ref{hybrid_schematic2} depicts the underlying concept of the hybrid CR scheme. At lower energies, the key states are fully fine-structure resolved, whereas higher-lying levels are grouped into superconfigurations. The continuum represents the ionization pathways and autoionizing states. Three categories of transitions are indicated as follows:

\begin{enumerate}

\item \textbf{\textit{S--S} transitions} connect one superconfiguration to another. These include transitions within singly excited superconfigurations (e.g., $(1)(6)$, $(1)(7)$, $(1)(8)$, etc.) and transitions from singly excited superconfigurations to doubly excited or autoionizing superconfigurations (e.g., $(2)^2$, $(2)(3)$, etc.). In the previous sentence, we express superconfigurations in the standard form $(n_1)^{N_1}(n_2)^{N_2}\dots (n_{\text{m}})^{N_{\text{m}}}$, where $n_m$ represents the principal quantum number of each orbital shell, and $N_m$ specifies the number of electrons occupying the corresponding shell. This notation corresponds to averaging over all configurations that arise from a Layzer complex~\cite{layzer1959}. If $N_m$ is omitted, it is assumed that $N_m = 1$.
Their effective cross-section, $\sigma^{SS}_{IF}$, is given by a statistical average over all the individual fine-structure levels  $i \in I$ within one superconfiguration $I$ and all levels $f \in F$ in the final superconfiguration $F$:
\begin{equation}
\sigma^{SS}_{IF} = 
\sum_{f \in F} \frac{\sum_{i \in I} g_{i}\,\sigma_{if}}{\sum_{i \in I} g_{i}},
\label{ss_cross}
\end{equation}
where $g_i$ is the statistical weight of the $i$th level and $\sigma_{if}$ is the fine-structure cross-section for the $i \rightarrow f$ transition.

\item \textbf{\textit{F--S} transitions} act as a ``handshake" between individually resolved fine-structure levels and superconfigurations. These transitions connect the lower resolved levels (e.g., $1s2s(^3S_1,^1S_0)$, $1s2p(^3P^o_{2,1,0},^1P^o_1)$, etc.) to upper superconfiguration blocks (e.g., $(1)(6)$, $(1)(7)$, $(2)^2$, $(2)(3)$, etc.). The total cross-section $\sigma^{FS}_{iF}$ for a fine-structure level $i$ transitioning into a superconfiguration $F$ is obtained by summing $\sigma_{if}$ over all final states $f$ in $F$:
\begin{equation}
\sigma^{FS}_{iF} = \sum_{f \in F} \sigma_{if}.
\end{equation}

\item \textbf{\textit{F--F} transitions} occur completely among fine-structure-resolved levels. Since each fine-structure level is individually retained, one simply uses the direct cross-sections (and rate coefficients) computed for these level-to-level transitions.

\end{enumerate}

As discussed earlier, the cross-sections and rate coefficients of the forward atomic processes, which include EIE, EII, PI, and AI, are taken directly from the fine-structure calculations produced by FAC. The quantities associated with the corresponding inverse processes, namely EIDE, 3BR, RR, and dielectronic recombination, are computed using the detailed balance relation and are applied uniformly across all transition classes (\textit{F–F}, \textit{F–S}, and \textit{S–S}). This procedure ensures that each forward and inverse cross-section or rate-coefficient pair is related through the appropriate statistical weights and Boltzmann factors, thereby maintaining the thermodynamic consistency of the hybrid CR model.

Notably, in the present work, the hybrid scheme is specifically implemented for singly excited states only, while doubly excited states are treated using the superconfiguration approach. For fusion-relevant electron densities, this selective treatment is justified, as the predominantly autoionizing nature of doubly-excited states prevents them from retaining significant population, which makes detailed fine-structure resolution less important than for the low-lying singly-excited states.

This partial-averaging approach, i.e., the hybrid scheme, significantly reduces the number of rate equations as well as the size of the rate matrix that needs to be solved, while still preserving the essential physics. Importantly, the fine-structure levels that play a dominant role in diagnostics and emission spectra remain $J$-resolved, whereas the superconfiguration approach captures the collective effect of higher-lying states. Hence, the hybrid scheme strikes a balance between accuracy and computational efficiency, enabling detailed CR modeling without requiring full fine-structure resolution across all excited levels.

\section{Results and Discussion}
In this work, we have developed two hybrid CR models, Hybrid-($n_\text{valence}+4$) and Hybrid-($n_\text{valence}$), for helium, lithium, and beryllium plasma species. The key distinction between these models lies in the amount of fine-structure resolution that is retained. For example, in the Hybrid-($n_\text{valence}+4$) model, fine-structure levels are retained up to $n_\text{valence}+4$, whereas higher-lying states with $n > n_\text{valence}+4$ (up to $n=10$) are grouped into superconfigurations. In contrast, the Hybrid-($n_\text{valence}$) model applies superconfiguration averaging starting at $n_\text{valence}+1$, offering a more compact representation at the cost of accuracy. A comprehensive list of fine-structure levels and superconfiguration states used in the Hybrid-($n_\text{valence}+4$) model for all three elements considered in this work is provided in Table~\ref{Table:config}.

To validate the hybrid scheme, we have benchmarked the hybrid CR models against a fully fine-structure CR model that resolves fine-structure levels up to $n=10$ for each species. Table~\ref{Table:hyb_level} lists the number of levels/states included in the fine-structure and hybrid models, respectively. We then calculated the key plasma properties, including the radiative power loss and average charge state, across electron densities from $10^{12}$ to $10^{17}$ cm$^{-3}$ and electron temperatures ranging from 1 to $10^{3}$ eV. The following sections present these comparisons in detail and demonstrate the fidelity of the proposed hybrid CR approach.

\begin{table}[htbp]
\centering
\caption{\label{Table:config} Fine-structure levels and superconfigurations for H-like, He-like, Li-like, and Be-like atoms/ions employed in the Hybrid-($n_\text{valence}+4$) CR model. Superconfigurations are expressed in the form $(n_1)^{N_1}(n_2)^{N_2}\dots (n_{\text{m}})^{N_{\text{m}}}$, where $n_m$ represents the principal quantum number of each orbital shell, and $N_m$ specifies the number of electrons occupying the corresponding shell.}

\setlength{\tabcolsep}{12pt}
\begin{tabular}{lllll}
\hline\hline
Shell & H-like& He-like & Li-like & Be-like \\
\hline
\multicolumn{5}{c}{Ground and Singly-excited states}\\
\hline
$n=1$ & $1s~(^2S_{1/2})$ & $1s^2~(^1S_0)$ & \multicolumn{1}{c}{--} & \multicolumn{1}{c}{--}\\
\hline
$n=2$ & $2p~(^2P^\circ_{1/2,3/2})$ & $1s~2s~(^3S_1,^1S_0)$  & $1s^2~2s~(^2S_{1/2})$ & $1s^2~2s^2~(^1S_0)$\\
& $2s~(^2S_{1/2})$ & $1s~2p~(^3P^o_{2,1,0},^1P^o_1)$  & \multicolumn{1}{c}{--} & $1s^2~2s~2p~(^3P^o_{0,1,2},^1P^o_1)$ \\
& \multicolumn{1}{c}{--}   & \multicolumn{1}{c}{--}  & \multicolumn{1}{c}{--} & $1s^2~2p^2~(^1D_2,^3P_{0,1,2})$\\
& \multicolumn{1}{c}{--}   & \multicolumn{1}{c}{--}  & \multicolumn{1}{c}{--}  & \multicolumn{1}{c}{--}\\
\hline
$n=3$  & $3p~(^2P^\circ_{1/2,3/2})$  &  $1s~3s(^3S_1,^1S_0)$  & $1s^2~3s(^2S_{1/2})$ & $1s^2~2s~3s~(^3S_1,^1S_0)$\\
& $3s~(^2S_{1/2})$  & $1s~3p(^3P^o_{0,1,2},^1P^o_1)$   & $1s^2~3p(^2P^o_{1/2,3/2})$ & $1s^2~2s~3p~(^3P^o_{0,1,2},^1P^o_1)$\\
& $3d~(^2D^\circ_{3/2,5/2})$  & $1s~3d(^3D_{1,2,3},^1D_2)$  & $1s^2~3d(^2D_{3/2,5/2})$ & $1s^2~2s~3d~(^3D_{1,2,3},^1D_2)$\\
\hline

$n=4$ & $4p~(^2P^\circ_{1/2,3/2})$  & $1s~4s(^3S_1,^1S_0)$      & $1s^2~4s(^2S_{1/2})$ & $1s^2~2s~4s~(^3S_1,^1S_0)$\\
    & $4s~(^2S_{1/2})$            & $1s~4p(^3P^o_{2,1,0},^1P^o_1)$ & $1s^2~4p(^2P^o_{1/2,3/2})$ & $1s^2~2s~4p~(^3P^o_{0,1,2},^1P^o_1)$\\
    & $4d~(^2D^\circ_{3/2,5/2})$  & $1s~4d(^3D_{1,2,3},^1D_2)$  & $1s^2~4d(^2D_{3/2,5/2})$ & $1s^2~2s~4d~(^3D_{1,2,3},^1D_2)$\\
    & $4f~(^2F^\circ_{5/2,7/2})$  & $1s~4f(^3F^o_{2,3,4},^1F^o_3)$  & $1s^2~4f(^2F^o_{5/2,7/2})$ & $1s^2~2s~4f~(^3F^o_{2,3,4},^1F^o_3)$\\
\hline
$n=5$ & $5p~(^2P^\circ_{1/2,3/2})$  & $1s~5s(^3S_1,^1S_0)$ & $1s^2~5s(^2S_{1/2})$ & $1s^2~2s~5s~(^3S_1,^1S_0)$\\
    & $5s~(^2S_{1/2})$            & $1s~5p(^3P^o_{2,1,0},^1P^o_1)$         & $1s^2~5p(^2P^o_{1/2,3/2})$ & $1s^2~2s~5p~(^3P^o_{0,1,2},^1P^o_1)$\\
    & $5d~(^2D^\circ_{3/2,5/2})$  & $1s~5d(^3D_{3,2,1},^1D_2)$   & $1s^2~5d(^2D_{3/2,5/2})$ & $1s^2~2s~5d~(^3D_{1,2,3},^1D_2)$\\
    & $5f~(^2F^\circ_{5/2,7/2})$  & $1s~5f(^3F^o_{4,3,2},^1F^o_3)$   & $1s^2~5f(^2F^o_{5/2,7/2})$ & $1s^2~2s~5f~(^3F^o_{2,3,4},^1F^o_3)$\\
    & $5f~(^2G^\circ_{7/2,9/2})$  & $1s~5g(^3G_{5,4,3},^1G_4)$         & $1s^2~5g(^2G^o_{7/2,9/2})$ & $1s^2~2s~5g~(^3G_{3,4,5},^1G_4)$\\
    \hline
\end{tabular}
\label{tab:atomic_states}
\end{table}

\begin{table}[htbp]
\centering
\setlength{\tabcolsep}{23pt}
\begin{tabular}{lllll}
\hline
$n=6$ 
& (6) & (1)~(6) & $1s^2~6s(^2S_{1/2})$ & $1s^2~2s~6s~(^3S_1,^1S_0)$\\
& \multicolumn{1}{c}{--} & \multicolumn{1}{c}{--} & $1s^2~6p(^2P^o_{1/2,3/2})$ & $1s^2~2s~6p~(^3P^o_{0,1,2},^1P^o_1)$\\
& \multicolumn{1}{c}{--} & \multicolumn{1}{c}{--} & $1s^2~6d(^2D_{3/2,5/2})$ & $1s^2~2s~6d~(^3D_{1,2,3},^1D_2)$\\
& \multicolumn{1}{c}{--} & \multicolumn{1}{c}{--} & $1s^2~6f(^2F^o_{5/2,7/2})$ & $1s^2~2s~6f~(^3F^o_{2,3,4},^1F^o_3)$\\
& \multicolumn{1}{c}{--} & \multicolumn{1}{c}{--} & $1s^2~6g(^2G^o_{7/2,9/2})$ & $1s^2~2s~6g~(^3G_{3,4,5},^1G_4)$\\
& \multicolumn{1}{c}{--} & \multicolumn{1}{c}{--} & $1s^2~6h(^2H^o_{9/2,11/2})$ & $1s^2~2s~6h~(^3H_{4,5,6},^1H^o_5)$\\
\hline
$n=7$ & (7) & (1)~(7) & (1)$^2$~(7) & (1)$^2$~(2)~(7)\\
\hline
$n=8$ & (8) & (1)~(8) & (1)$^2$~(8) & (1)$^2$~(2)~(8)\\
\hline
$n=9$ & (9) & (1)~(9) & (1)$^2$~(9) & (1)$^2$~(2)~(9)\\
\hline
$n=10$ & (10) & (1)~(10) & (1)$^2$~(10) & (1)$^2$~(2)~(10)\\
\hline\hline
\multicolumn{5}{c}{Doubly-excited states}\\
\hline
\hline
$n=2$ & \multicolumn{1}{c}{--} & (2)$^2$ & (1)~(2)$^2$ & (1)~(2)$^3$\\
\hline
$n=3$ & \multicolumn{1}{c}{--} & (2)~(3) & (1)~(2)~(3) & (1)~(2)$^2$(3),~(1)$^2$~(3)$^2$\\
\hline
$n=4$ & \multicolumn{1}{c}{--} & (2)~(4) & (1)~(2)~(4) & (1)~(2)$^2$(4),~(1)$^2$~(3)~(4)\\
\hline
$n=5$ & \multicolumn{1}{c}{--} & (2)~(5) & (1)~(2)~(5) & (1)~(2)$^2$(5),~(1)$^2$~(3)~(5)\\
\hline
$n=6$ & \multicolumn{1}{c}{--} & (2)~(6) & (1)~(2)~(6) & (1)~(2)$^2$(6),~(1)$^2$~(3)~(6)\\
\hline
$n=7$ & \multicolumn{1}{c}{--} & (2)~(7) & (1)~(2)~(7) & (1)~(2)$^2$(7),~(1)$^2$~(3)~(7)\\
\hline
$n=8$ & \multicolumn{1}{c}{--} & (2)~(8) & (1)~(2)~(8) & (1)~(2)$^2$(8),~(1)$^2$~(3)~(8)\\
\hline
$n=9$ & \multicolumn{1}{c}{--} & (2)~(9) & (1)~(2)~(9) & (1)~(2)$^2$(9),~(1)$^2$~(3)~(9)\\
\hline
$n=10$ & \multicolumn{1}{c}{--} & (2)~(10) & (1)~(2)~(10) & (1)~(2)$^2$(10),~(1)$^2$~(3)~(10)\\
\hline
\hline
\end{tabular}
\label{tab:atomic_states}
\end{table}

\begin{table}[!h]
\centering
\setlength{\tabcolsep}{28pt}
\caption{\label{Table:hyb_level} Comparison of the number of levels/states in the fine-structure, Hybrid-($n_{\text{valence}}+4$), and Hybrid-($n_{\text{valence}}$) CR models for different elements.}
\begin{tabular}{cccc}
\hline
\hline
&Fine-structure & Hybrid-($n_\text{valence}$+4) & Hybrid-($n_\text{valence}$) \\
\hline
Helium      & 1046        & 94      & 30  \\
Lithium      & 2585        & 142      & 50  \\
Beryllium      & 9790        & 413      & 85  \\
\hline
\hline               
\end{tabular}

\end{table}

\clearpage
\subsection{Helium}
Helium plays a central role in fusion plasmas, serving not only as the primary fusion by-product in D–T reactions but also as a commonly employed plasma-conditioning and diagnostic gas in experimental plasma operations. In the context of CR modeling, the recombination and ionization processes are crucially important for helium ash extraction or recycling~\cite{Reiter2004} in particular.
Its two-electron structure produces rich atomic kinetics with strong electron–electron correlations, singlet–triplet splitting with long-lived metastables, and numerous doubly excited states that autoionize and enable dielectronic recombination.

\begin{figure}[!h]
\centering
\includegraphics[scale=0.25]{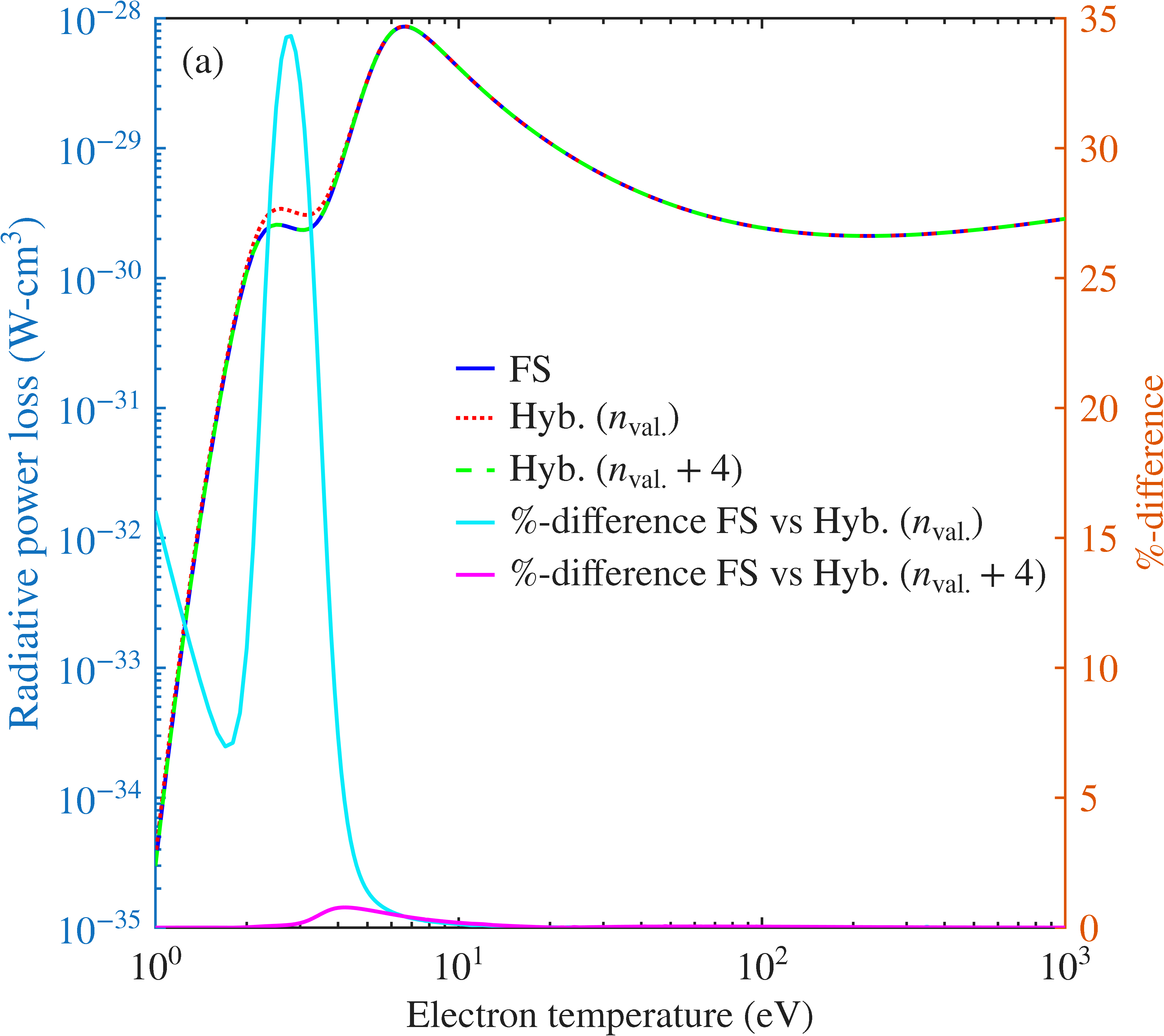} \hspace{2mm} \includegraphics[scale=0.25]{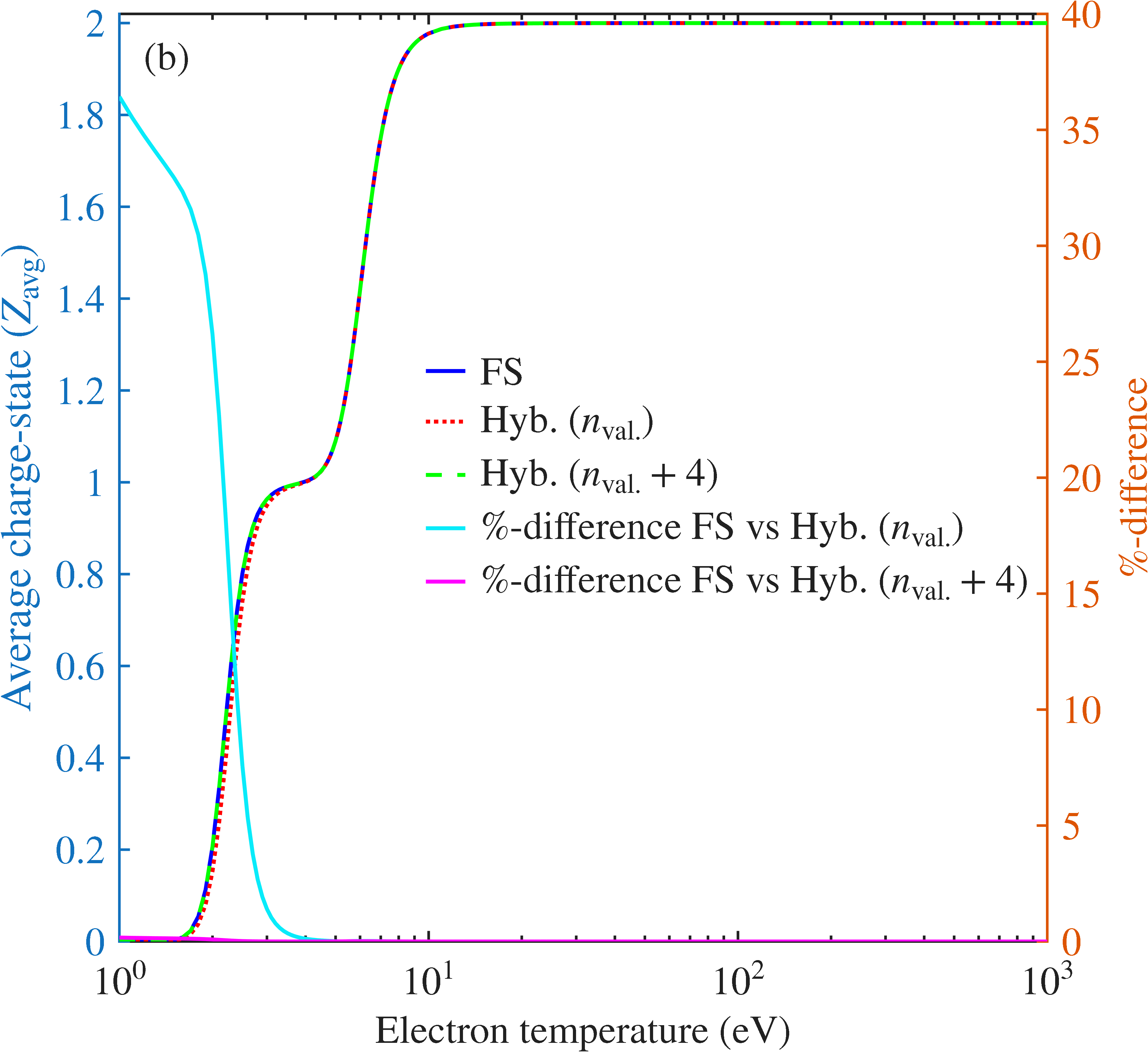}
\includegraphics[scale=0.25]{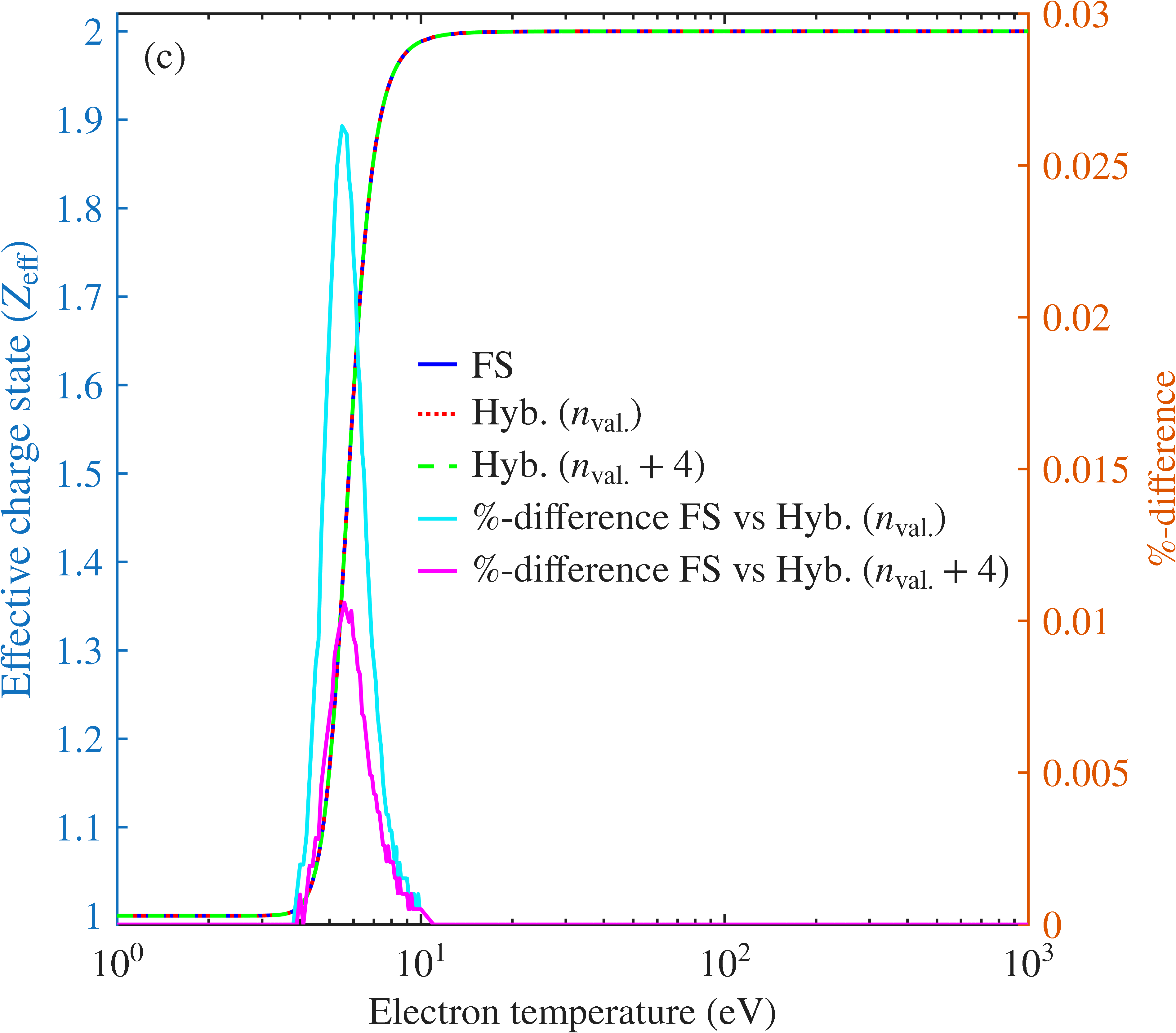}
\caption{\label{Hybrid_Compare_He} 
\textbf{Helium:} Comparison of (a) radiative power loss, (b) average charge state, and (c) effective charge state obtained from the fine-structure CR model and two hybrid models, Hybrid-($n_\text{valence}$) and Hybrid-($n_\text{valence}+4$), at $n_e = 10^{14}$ cm$^{-3}$. The right-hand axes show the percentage deviation of each hybrid model from the fine-structure results. Here, "val." = valence and "Hyb." = Hybrid.}
\end{figure}

Figure~\ref{Hybrid_Compare_He} shows a comparison of the radiative power loss, average charge state, and effective charge state at a fusion-relevant electron density of $10^{14}~\text{cm}^{-3}$, representative of edge and divertor plasma conditions. Both hybrid models follow the fine-structure benchmark over most of the temperature range; however, the Hybrid-($n_{\text{valence}}$) model shows significant deviations at lower temperatures. In the peak radiative window ranging from 5--50~eV, the Hybrid-($n_{\text{valence}}+4$) model consistently remains within 2\% of the fine-structure results for all quantities, whereas the reduced model exhibits errors up to 35\% near 3--4~eV. This temperature range corresponds to the transition from neutral to singly ionized helium, where limited level resolution introduces averaging errors in spontaneous radiative decay rates and leads to deviations in the predicted radiative power loss.

At electron temperatures above 20~eV, both hybrid models converge to within 1\% of the fine-structure benchmark. This behavior reflects the fact that the plasma is largely ionized in this regime, and the contribution of low-lying fine-structure states to the radiative balance becomes minor.

\begin{figure}[!h]
\centering
\includegraphics[scale=0.25]{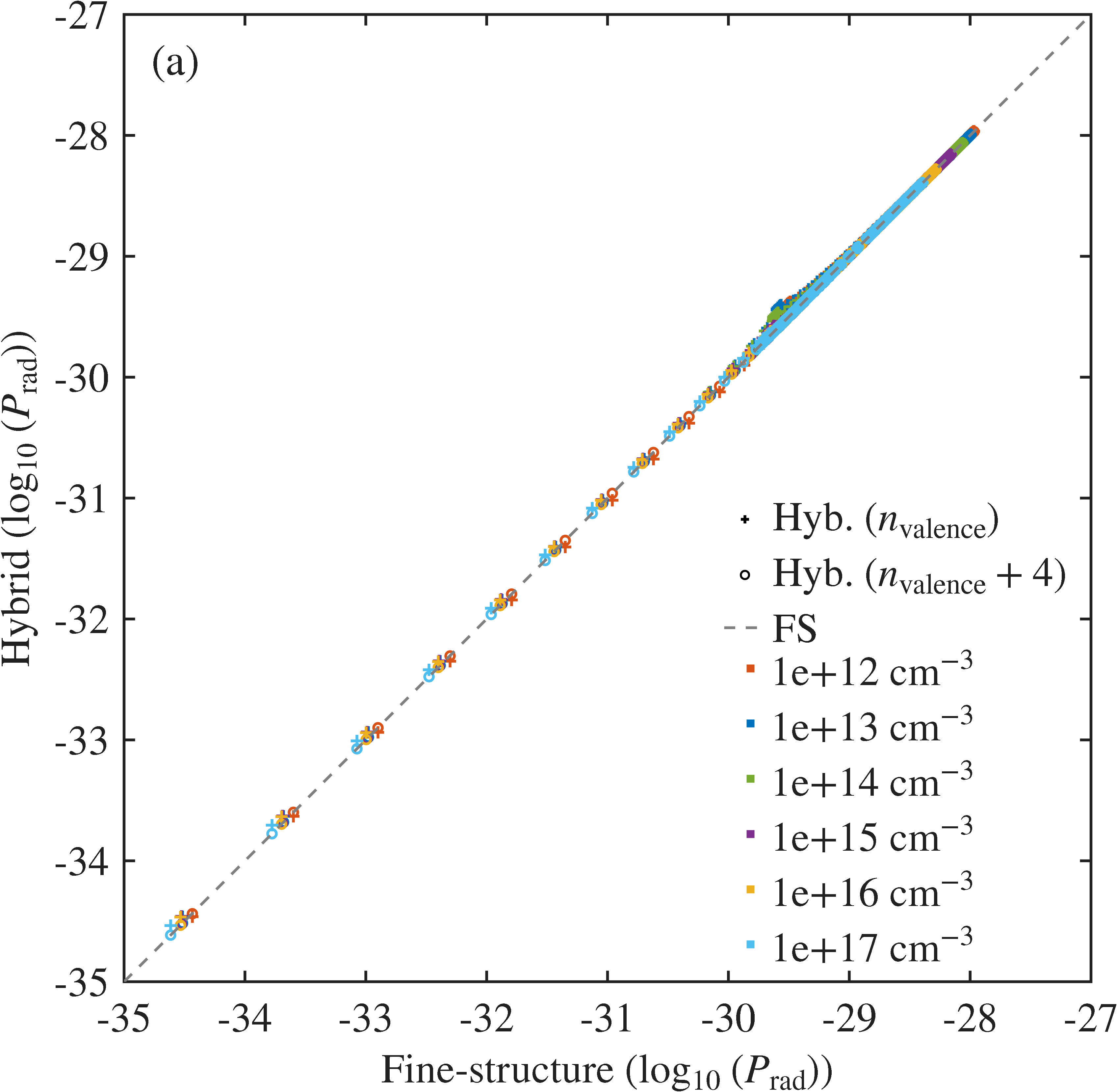} \hspace{2mm} \includegraphics[scale=0.25]{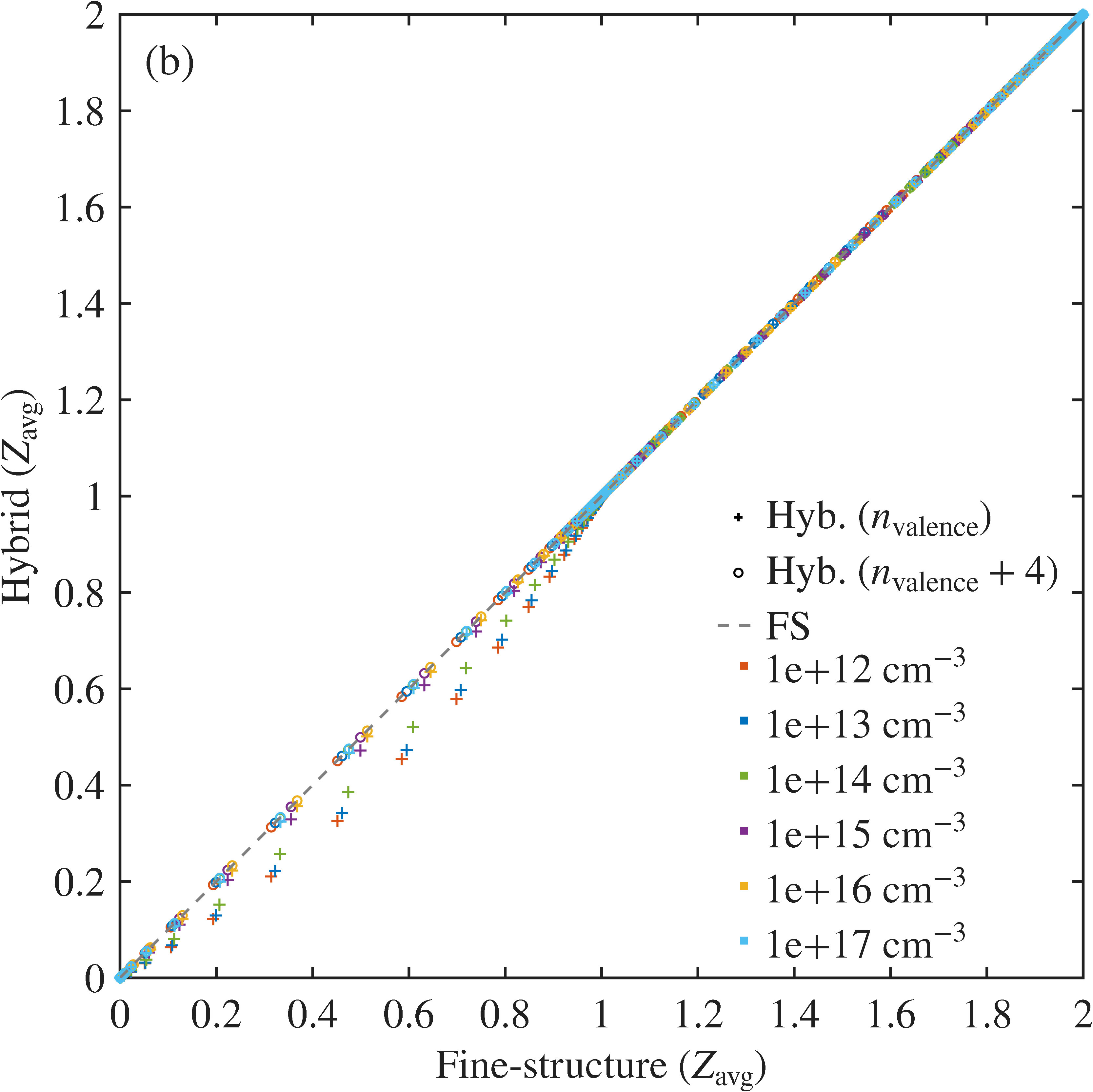}
\includegraphics[scale=0.25]{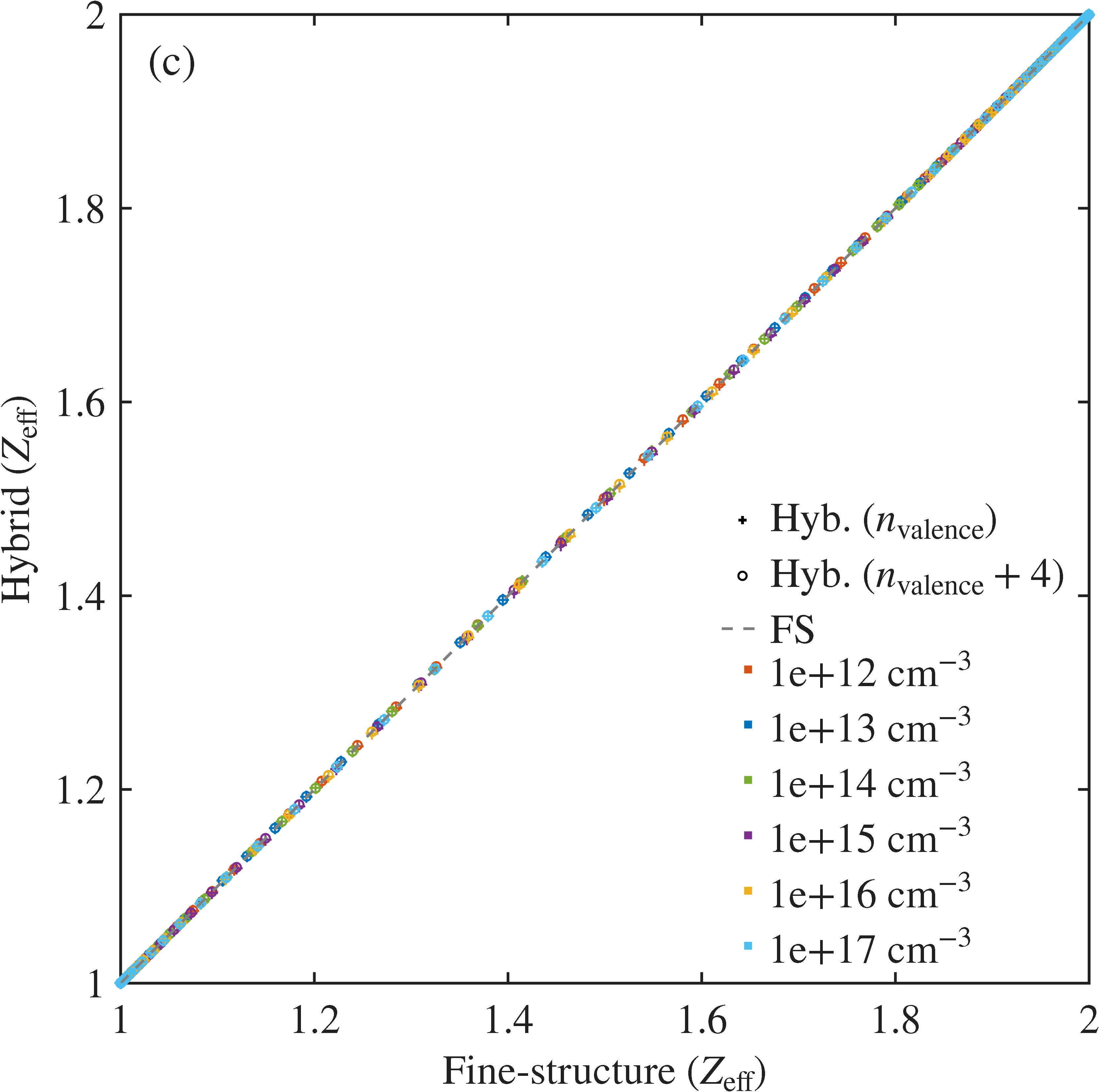}
\caption{\label{Hyb_Fine_Parity_He}
\textbf{Helium:} Parity plots comparing (a) radiative power loss, $P_\mathrm{rad}$, (b) average charge state, $Z_\mathrm{avg}$, and (c) effective charge state, $Z_\mathrm{eff}$, predicted by the Hybrid-($n_\text{valence}$) and Hybrid-($n_\text{valence}+4$) models against the fine-structure CR model across electron densities from $10^{12}$ to $10^{17}$~cm$^{-3}$. The dashed diagonal represents perfect agreement ($y = x$); points closer to the line indicate better agreement with the fine-structure benchmark.}

\end{figure}

To further support these findings, we have included parity plots, as shown in Fig.~\ref{Hyb_Fine_Parity_He}, which provide a direct visual comparison between the hybrid predictions and the fine-structure benchmark. In these plots, each point represents a pair of values from the two models, and perfect agreement places the point on the diagonal line. Points clustered near the diagonal indicate that the Hybrid-($n_{\text{valence}}+4$) model successfully reproduces the fine-structure results with high accuracy, while noticeable deviations reveal the limitations of the reduced Hybrid-($n_{\text{valence}}$) model. These discrepancies become more significant at lower electron densities, where metastable states are not collisionally quenched and can retain larger populations. Their longer lifetimes and density-sensitive transition pathways make the atomic kinetics highly sensitive to level resolution. As the reduced Hybrid-($n_{\text{valence}}$) model does not fully capture these metastable contributions, it exhibits larger deviations from the benchmark fine-structure results in this regime.

\clearpage
\subsection{Lithium}
In fusion plasmas, lithium is introduced both as an evaporated wall-conditioning layer~\cite{kaita2019fusion} and as a candidate liquid-metal plasma-facing component~\cite{mirnov2009plasma}, thereby entering the plasma as a controlled impurity and diagnostic species in many devices. With three electrons, lithium exhibits richer atomic physics than helium: low-lying open-shell structure, numerous doubly excited states, and dense autoionizing and dielectronic recombination resonances provide additional pathways for atomic kinetics and radiation emission. 

\begin{figure}[!h]
\centering
\includegraphics[scale=0.25]{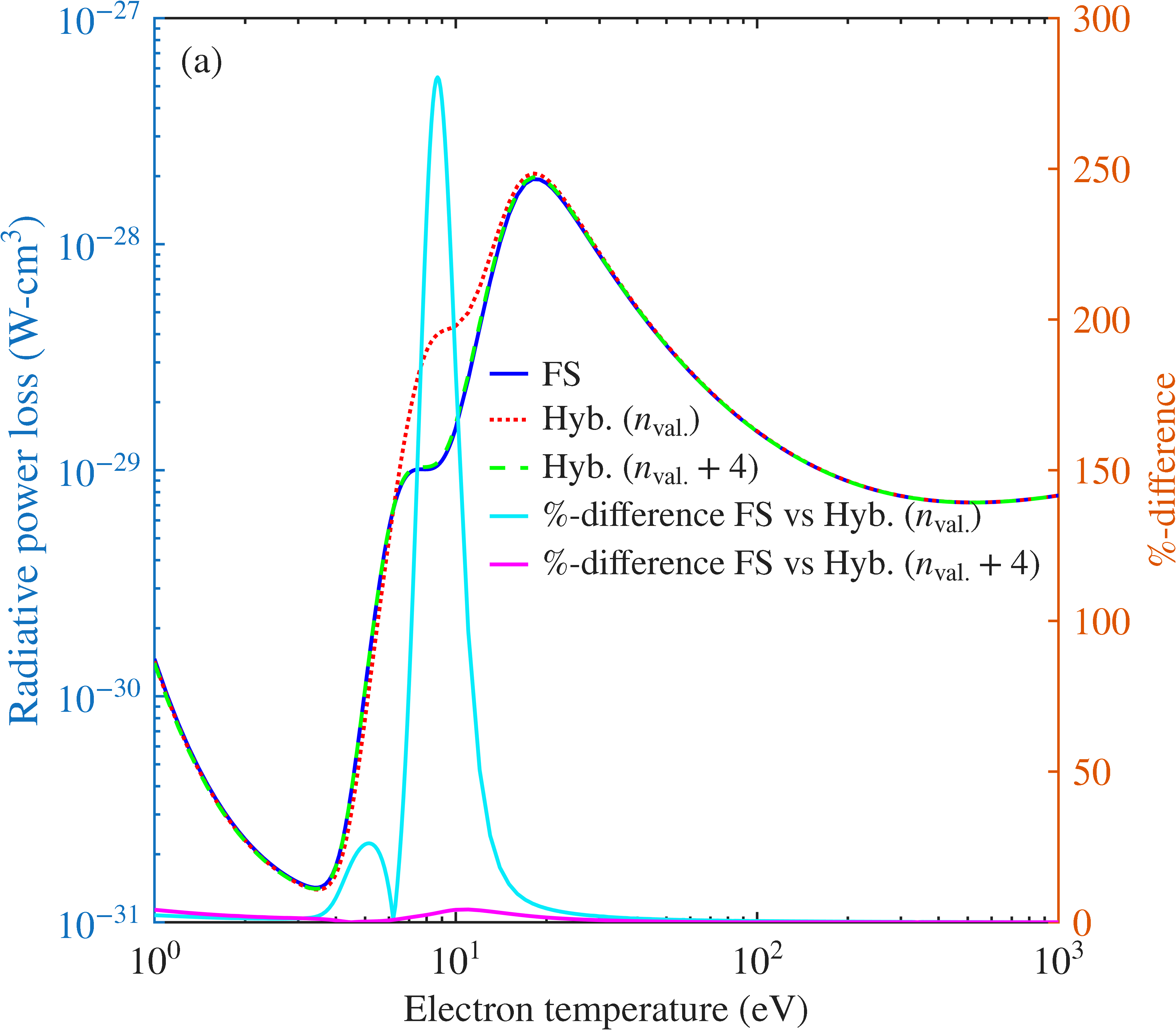} \hspace{2mm} \includegraphics[scale=0.25]{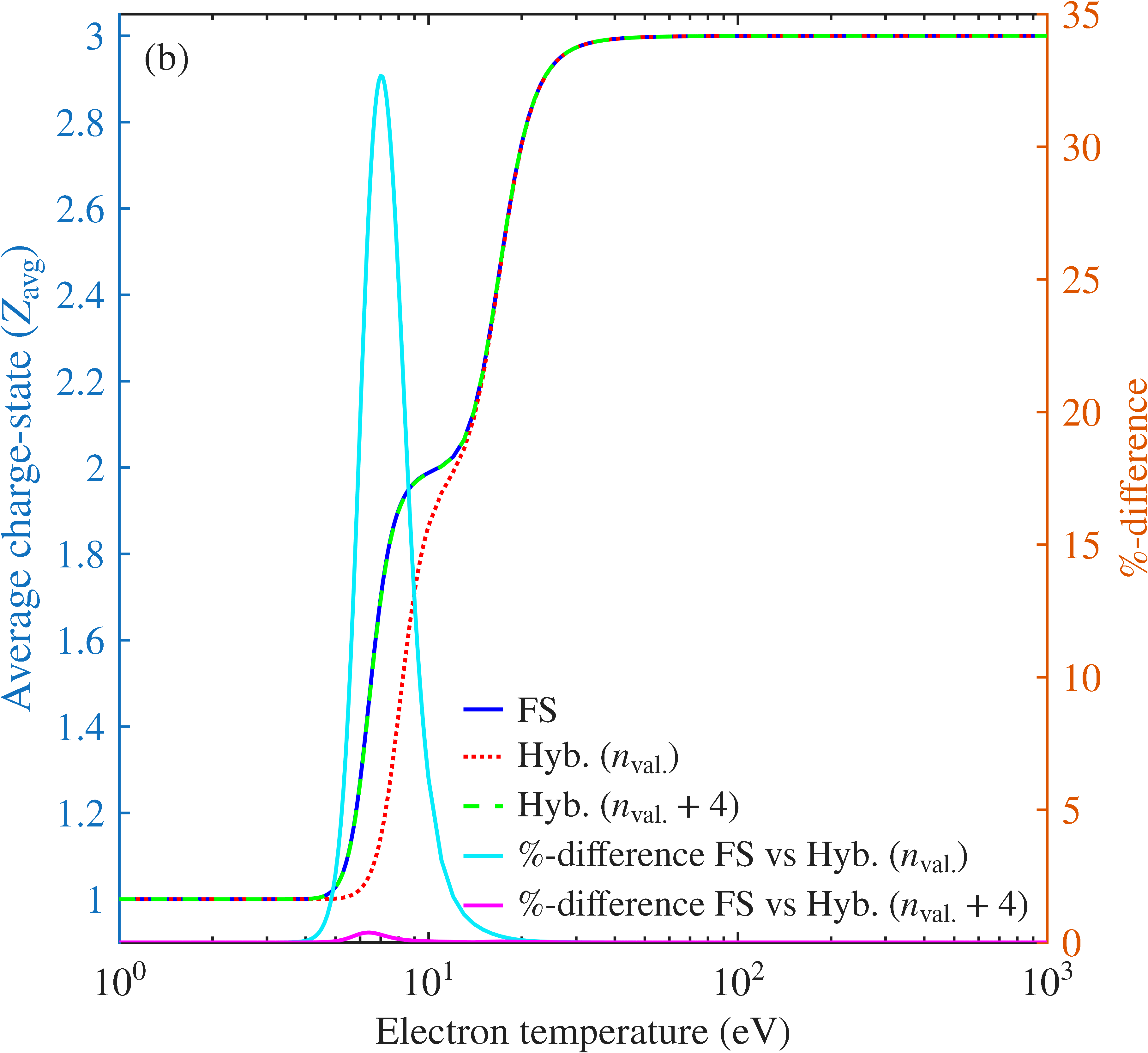}
\includegraphics[scale=0.25]{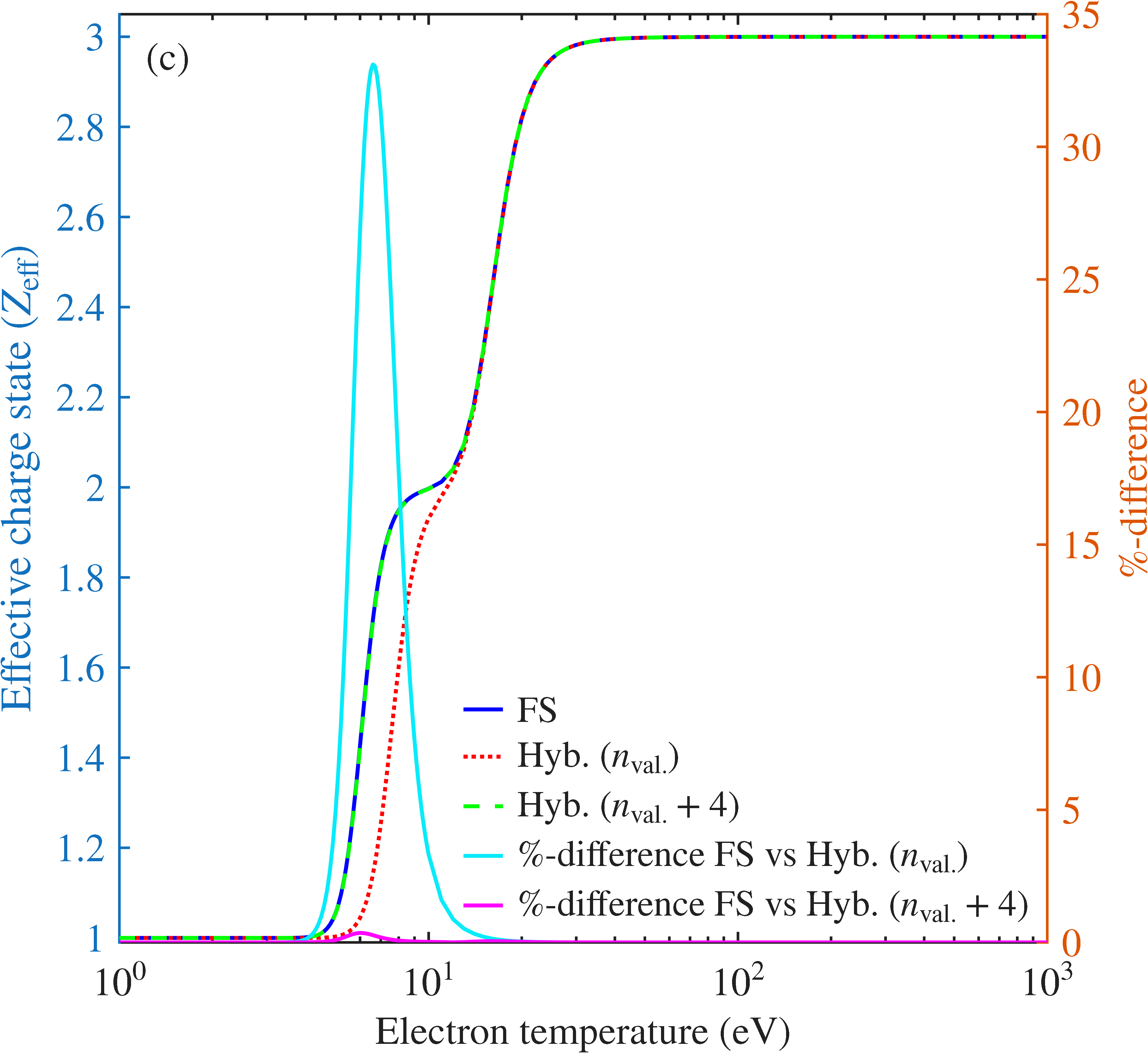}
\caption{\label{Hybrid_Compare_Li} \textbf{Lithium:} Comparison of (a) radiative power loss, (b) average charge state, and (c) effective charge state obtained from the fine-structure CR model and two hybrid models, Hybrid-($n_\text{valence}$) and Hybrid-($n_\text{valence}+4$), at $n_e = 10^{14}$~cm$^{-3}$. Right-hand axes show the percentage deviation of each hybrid model from the fine-structure results. Here, "val." = valence and "Hyb." = Hybrid.}
\end{figure}

Figure \ref{Hybrid_Compare_Li} presents a comprehensive comparison at $n_e = 10^{14}$~cm$^{-3}$ for all three key plasma properties, revealing the distinctly different behavior of lithium compared to helium. The radiative power loss exhibits three distinct maxima with a low-temperature peak near $T_e \approx 1$~eV associated with neutral Li line radiation, a sharp dominant peak near $T_e \approx 8$~eV arising from strong Li$^{+}$ emission, and a broader maximum around $T_e \approx 25$~eV corresponding to Li$^{2+}$. This triple-peak structure reflects the sequential ionization of the three electrons of Li. The Hybrid-($n_{\text{valence}} + 4$) model captures this complex radiative structure with excellent accuracy, closely reproducing both the sharp primary peak and the broader secondary maximum. The Hybrid-($n_{\text{valence}}$) model, however, significantly underestimates the secondary radiative peak, demonstrating the critical importance of retaining fine-structure details for higher-lying states in the hybrid treatment.

The ionization balance presents an even more demanding test, exhibiting three well-separated plateaus corresponding to the successive Li $\rightarrow$ Li$^+$ $\rightarrow$ Li$^{2+}$ $\rightarrow$ Li$^{3+}$ ionization stages. Each transition occurs over a relatively narrow temperature window, creating sharp increases in the average charge-state curve. The Hybrid-($n_{\text{valence}} + 4$) model reproduces these ionization transitions to within 1\% accuracy across the full temperature range, capturing both the plateau regions and the sharp transition zones with high fidelity. The reduced hybrid model ($n_{\text{valence}}$), however, systematically underestimates the charge state in the critical 4--20~eV window, where the Li$^+$ $\rightarrow$ Li$^{2+}$ transition dominates the plasma kinetics. This behavior arises from the complex interplay between ionization and recombination processes in the three-electron system, where refinement of singly excited states significantly alters the ionization balance. Further, although both models treat doubly excited states identically, another important consideration is that their different coupling with singly excited states can alter the ionization balance through the competition between autoionization and spontaneous decay from the doubly excited configurations.

The effective charge state exhibits similar structural features at the 4--20~eV electron temperature. The extended hybrid model maintains a robust performance with deviations below 1\% throughout this rapidly varying region, whereas the reduced hybrid model shows 30--35\% discrepancies.

\begin{figure}[!h]
\centering
\includegraphics[scale=0.25]{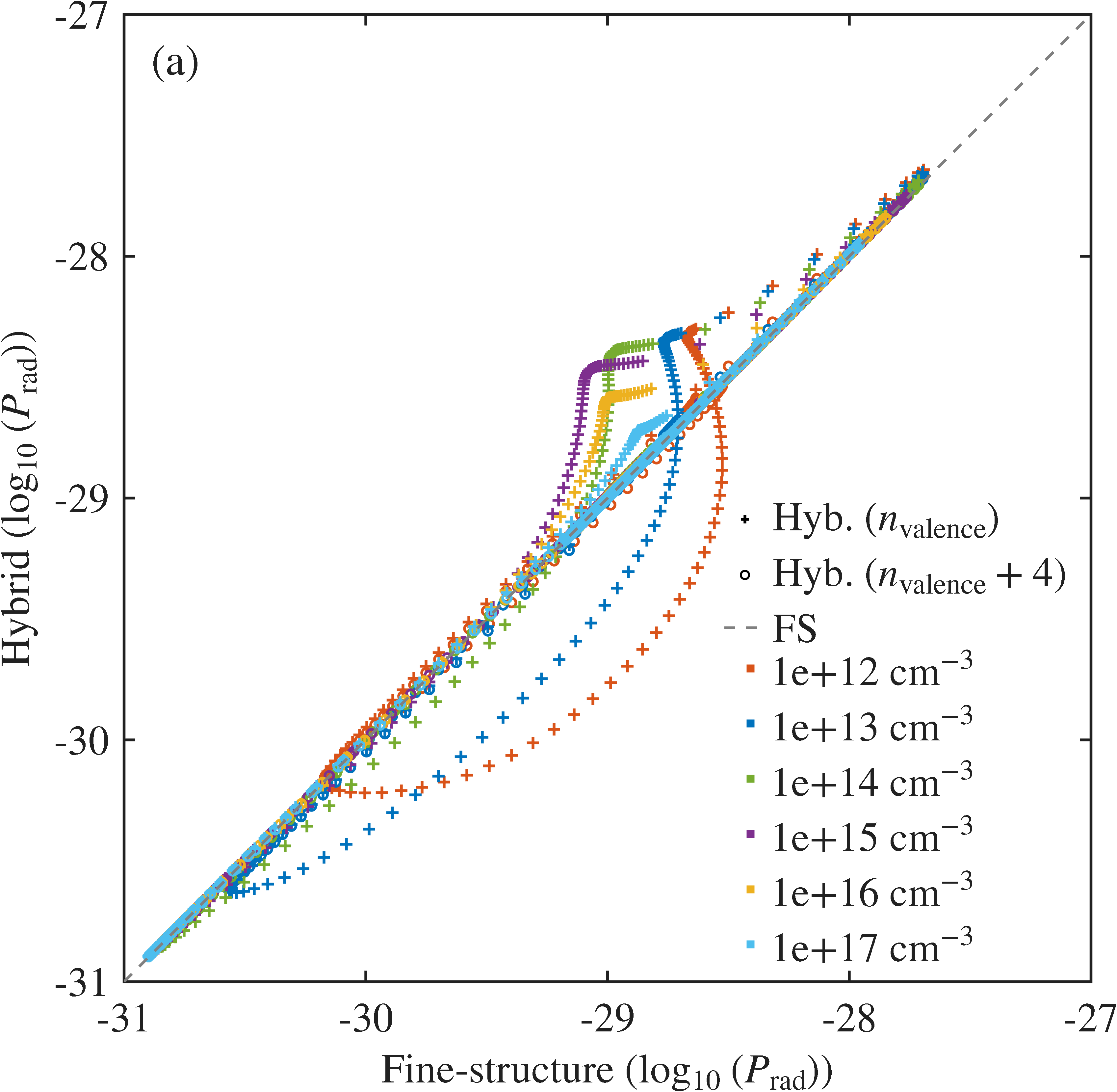} \hspace{2mm} \includegraphics[scale=0.25]{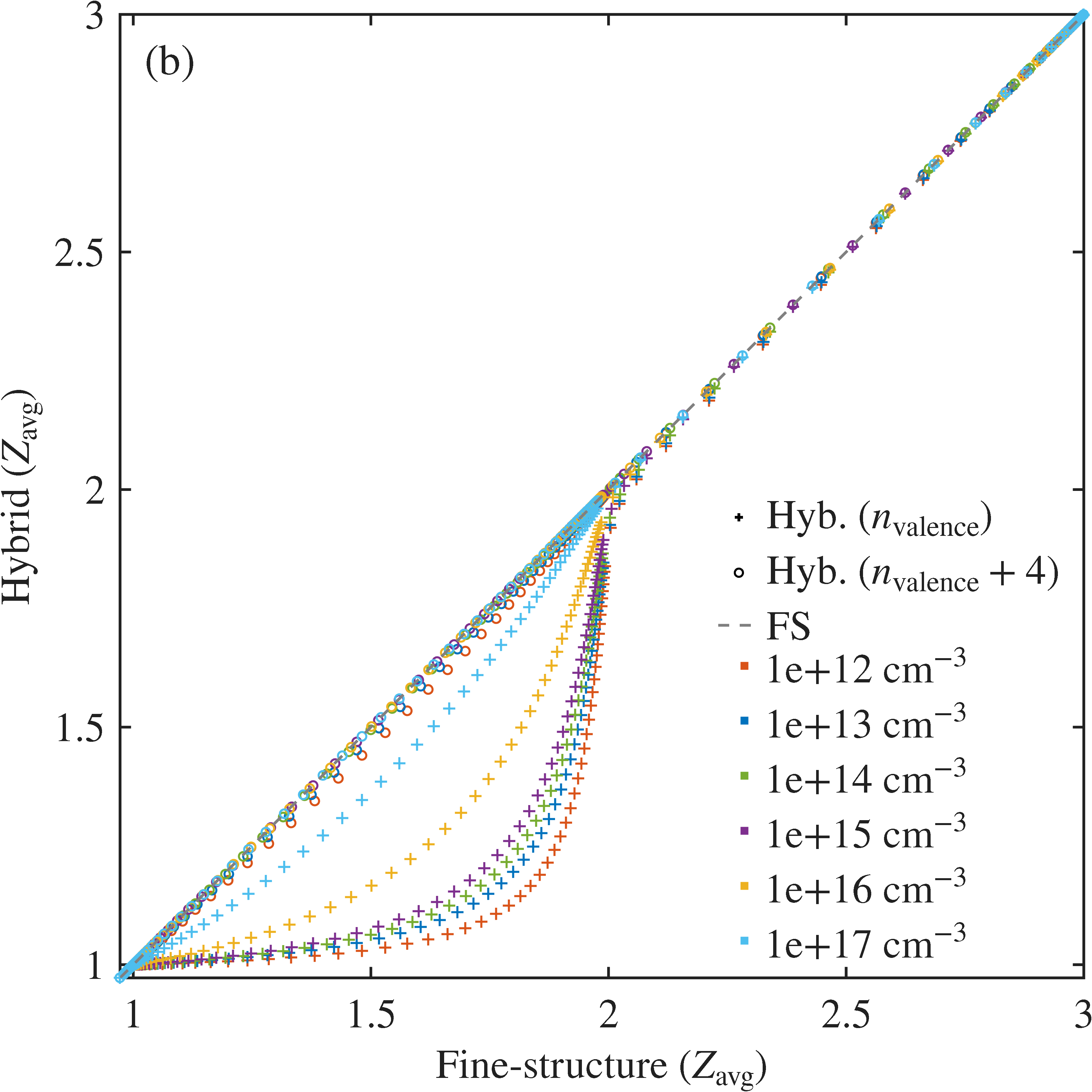}
\includegraphics[scale=0.25]{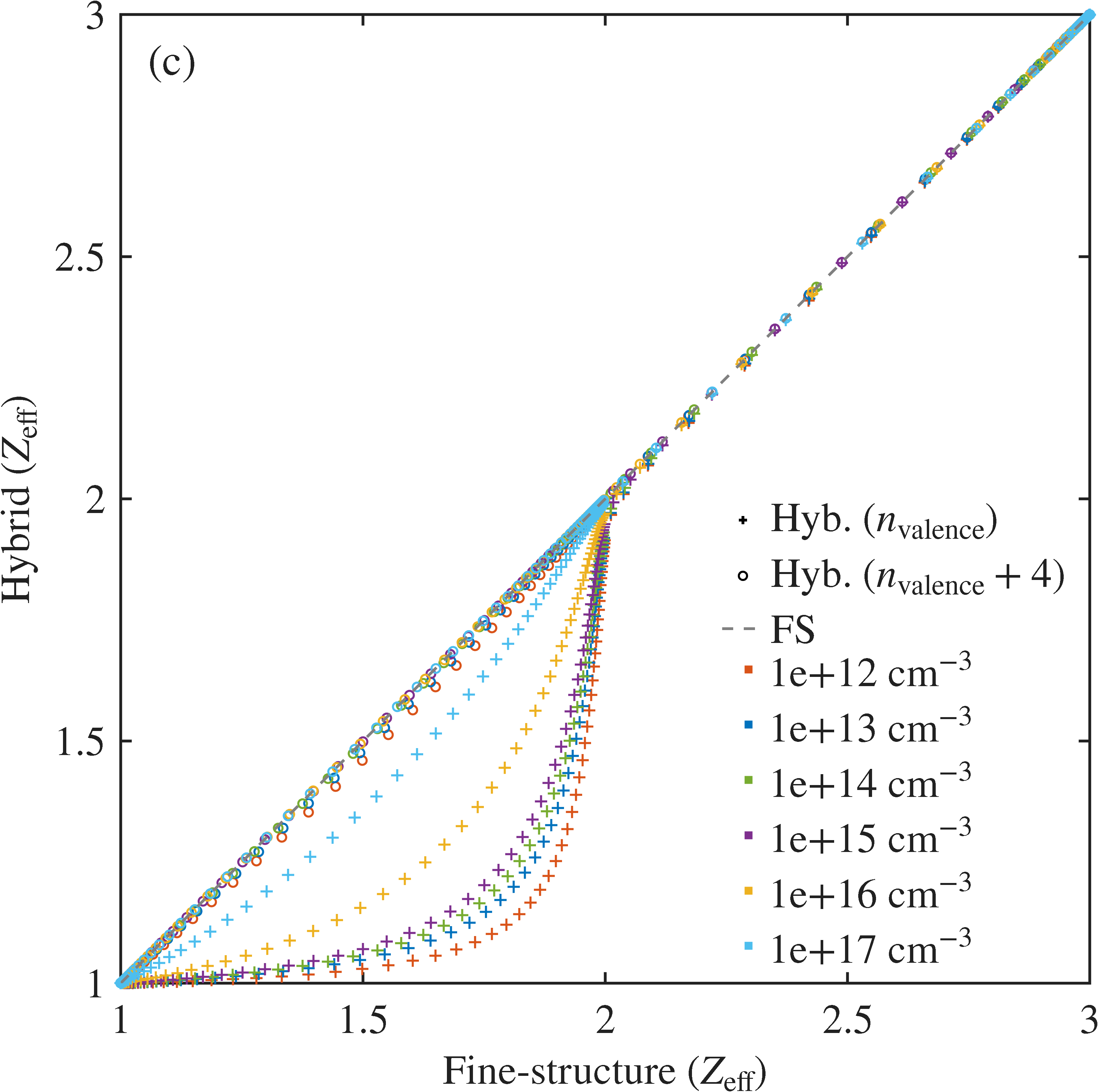}
\caption{\label{Hyb_Fine_Parity_Li}
\textbf{Lithium:} Parity plots comparing (a) radiative power loss, $P_\mathrm{rad}$, (b) average charge state, $Z_\mathrm{avg}$, and (c) effective charge state, $Z_\mathrm{eff}$, predicted by Hybrid-($n_\text{valence}$) and Hybrid-($n_\text{valence}+4$) models against the fine-structure CR model across electron densities from $10^{12}$ to $10^{17}$~cm$^{-3}$. The dashed diagonal represents perfect agreement ($y = x$); points closer to the line indicate better agreement with the fine-structure benchmark.}

\end{figure}

The parity plots in Fig.~\ref{Hyb_Fine_Parity_Li} provide quantitative confirmation that lithium demands substantially finer resolution than helium to achieve comparable accuracy. While the Hybrid-($n_{\text{valence}} + 4$) results maintain a near-diagonal alignment across the full parameter space, the reduced hybrid model exhibits significant scatter, particularly for radiative power-loss predictions. These results establish that for three-electron systems, such as lithium, the extended hybrid approach is essential for reliable plasma modeling.

\clearpage
\subsection{Beryllium}

Beryllium is another useful benchmark for fusion plasmas as it has
been a plasma-facing material on the ITER-like first wall on
JET~\cite{Rubel2018} and thus an important impurity that must be
understood for interpreting JET plasma scenarios.  Although the
updated design of ITER has opted for a full tungsten first wall in
lieu of the original beryllium first wall design, beryllium remains an
interesting first wall material for its low atomic number and being an
excellent neutron multiplier.  Its four-electron structure (Be I:
$1s^2 2s^2$) introduces multi-electron correlations, doubly excited
states, and dielectronic recombination processes, making the atomic
kinetics more complex than those of lithium while remaining tractable
in the fine-structure description, which provides the benchmark to
assess the physics fidelity of the hybrid models in the present FCR
code.

\begin{figure}[!h]
\centering
\includegraphics[scale=0.25]{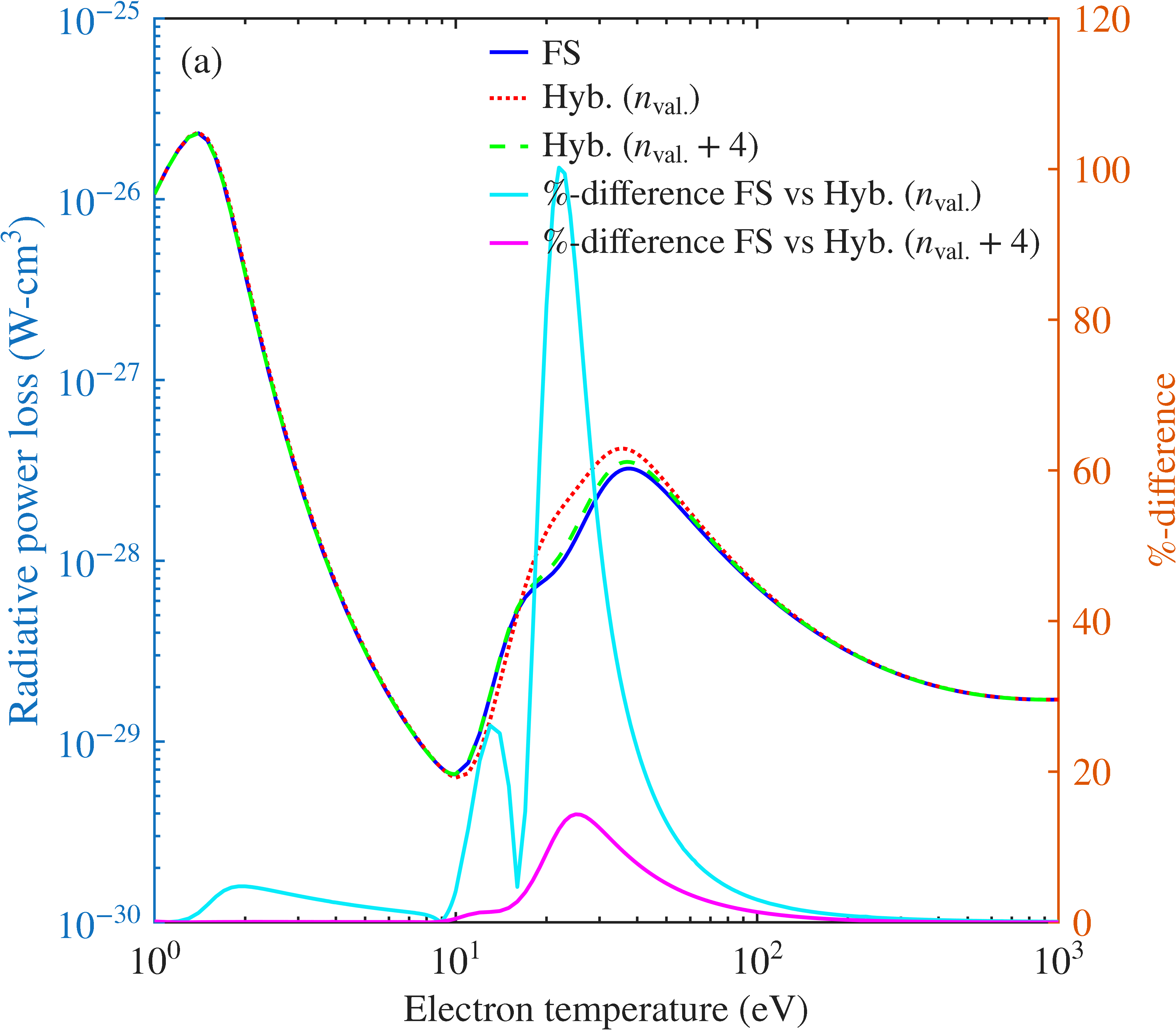} \hspace{2mm} \includegraphics[scale=0.25]{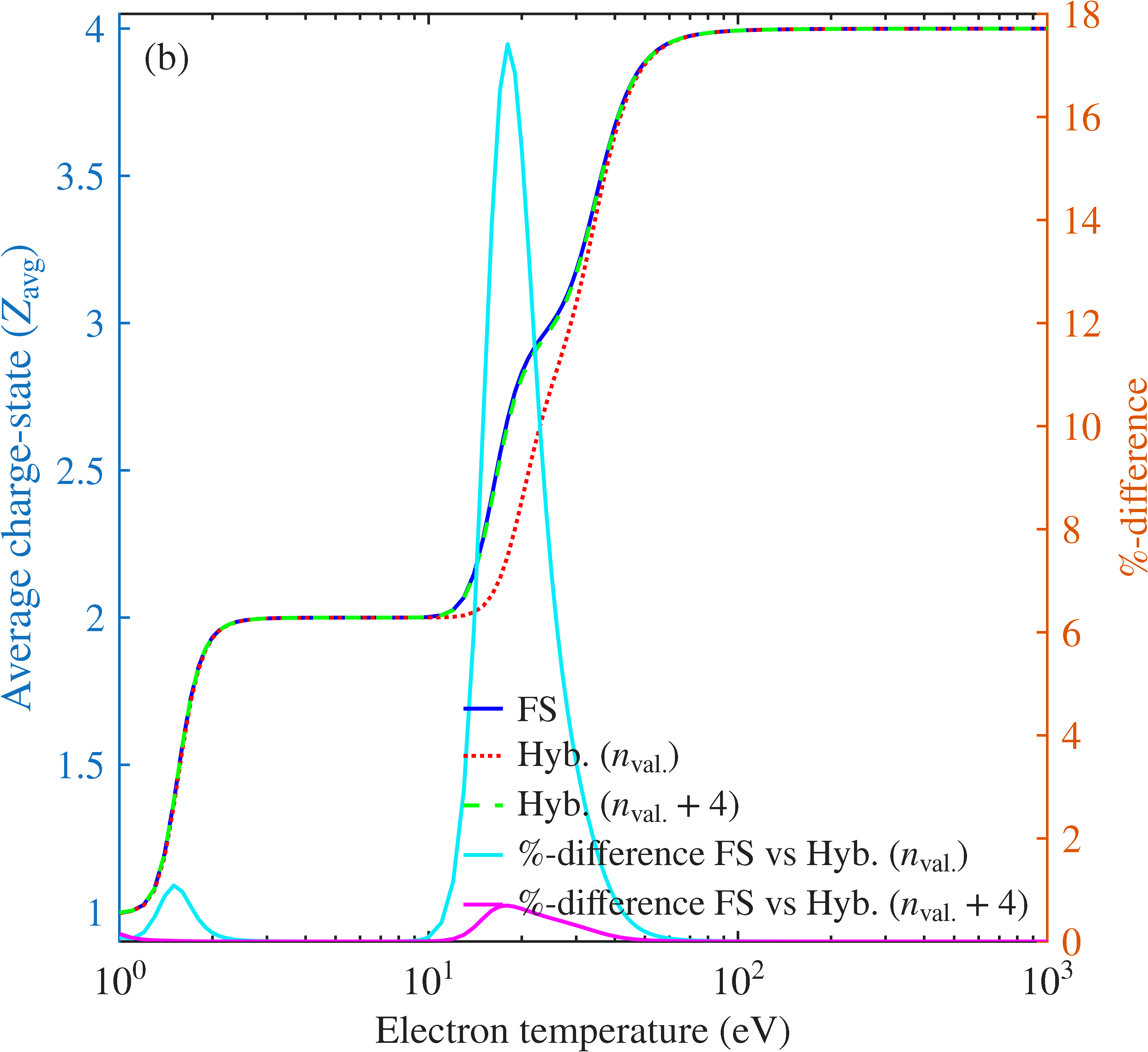}
\includegraphics[scale=0.25]{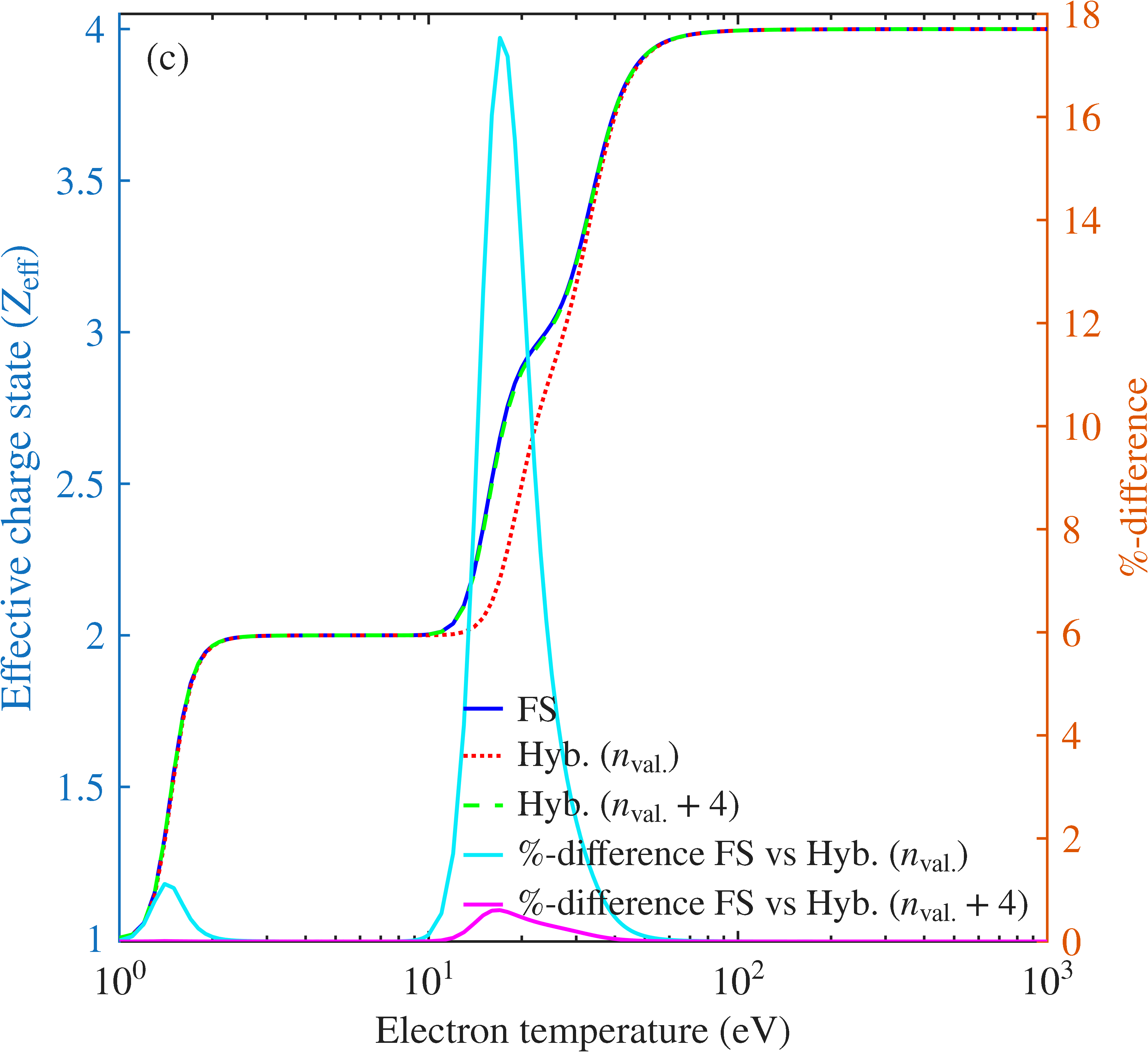}
\caption{\label{Hybrid_Compare_Be} \textbf{Beryllium:} Comparison of (a) radiative power loss, (b) average charge state, and (c) effective charge state obtained from the fine-structure CR model and two hybrid models, Hybrid-($n_\text{valence}$) and Hybrid-($n_\text{valence}+4$), at $n_e = 10^{14}$~cm$^{-3}$. Right-hand axes show the percentage deviation of each hybrid model from the fine-structure results. Here, "val." = valence and "Hyb." = Hybrid.}
\end{figure}

At $n_e = 10^{14}$~cm$^{-3}$, Fig.~\ref{Hybrid_Compare_Be} shows that beryllium 
exhibits the most complex behavior among the three elements studied across all 
plasma properties. The radiative power loss displays a rich multi-peak structure, 
with primary, secondary, and tertiary maxima near 1--2~eV, 15--20~eV, and 30--40~eV, 
respectively. This structure arises from the sequential ionization from Be to 
Be$^{+}$, Be$^{2+}$, Be$^{3+}$, and Be$^{4+}$, with each ionization stage 
contributing distinct spectroscopic features dominated by different atomic 
transitions.

The Hybrid-($n_{\text{valence}}+4$) model reproduces this complex
radiative structure with deviations below 15\% across all temperature
ranges and peaks.  This level of accuracy is notable given the
computational reduction achieved, as the number of levels decreases
from nearly ten thousand in the fine-structure model to 413 in the
extended hybrid model. The Hybrid-($n_{\text{valence}}$) model fails
to capture significant fine-structure contributions, with deviations
of about 10\% at the first radiative peak and increasing to
approximately 30\% and 100\% at subsequent peaks, highlighting the
importance of retaining fine-structure detail for higher-lying states
in multi-electron systems.

The charge-state evolution exhibits four distinct ionization plateaus, each 
corresponding to a different beryllium ion. The transitions between these 
plateaus are sharper than those observed in lithium, reflecting more tightly 
bound electron configurations and steeper ionization potential gradients. The 
Hybrid-($n_{\text{valence}}+4$) model accurately captures these ionization 
dynamics with deviations below 2\% across the entire temperature range. The 
reduced hybrid model shows larger errors in critical transition regions, 
particularly during the Be$^{+}$ to Be$^{2+}$ transition, where many-body effects 
dominate the population kinetics.

\begin{figure}[!h]
\centering
\includegraphics[scale=0.25]{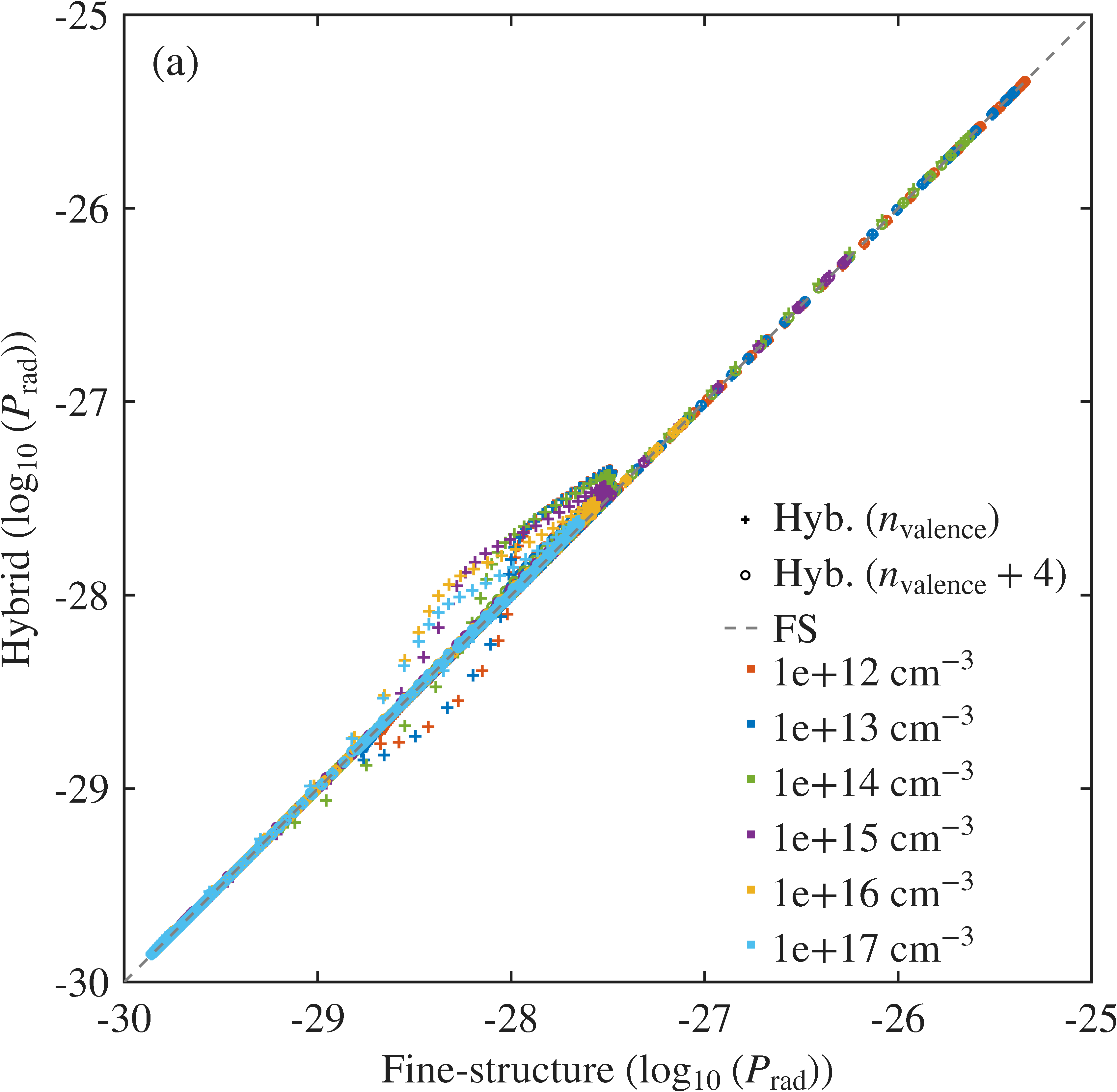} \hspace{2mm} \includegraphics[scale=0.25]{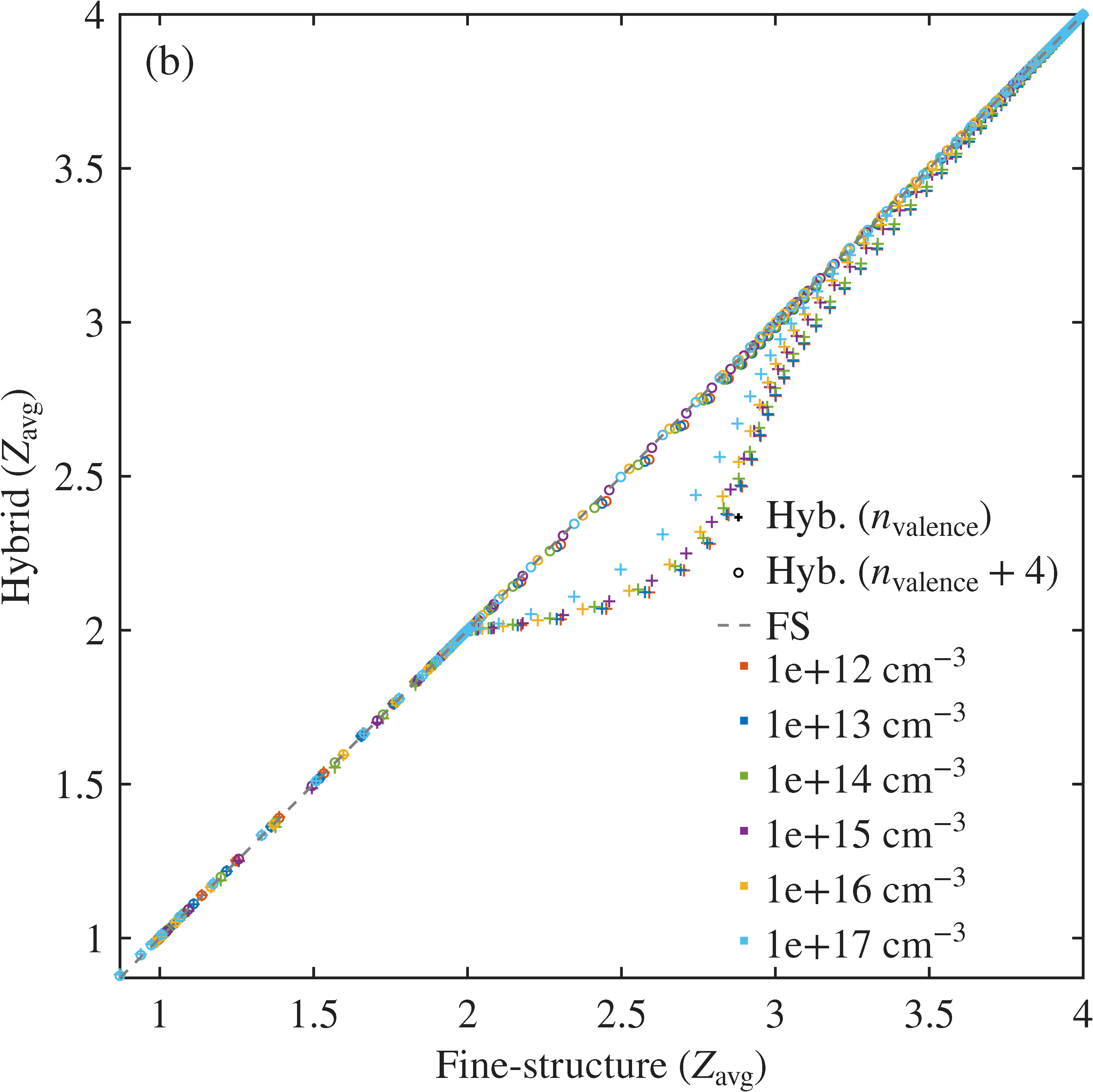}
\includegraphics[scale=0.25]{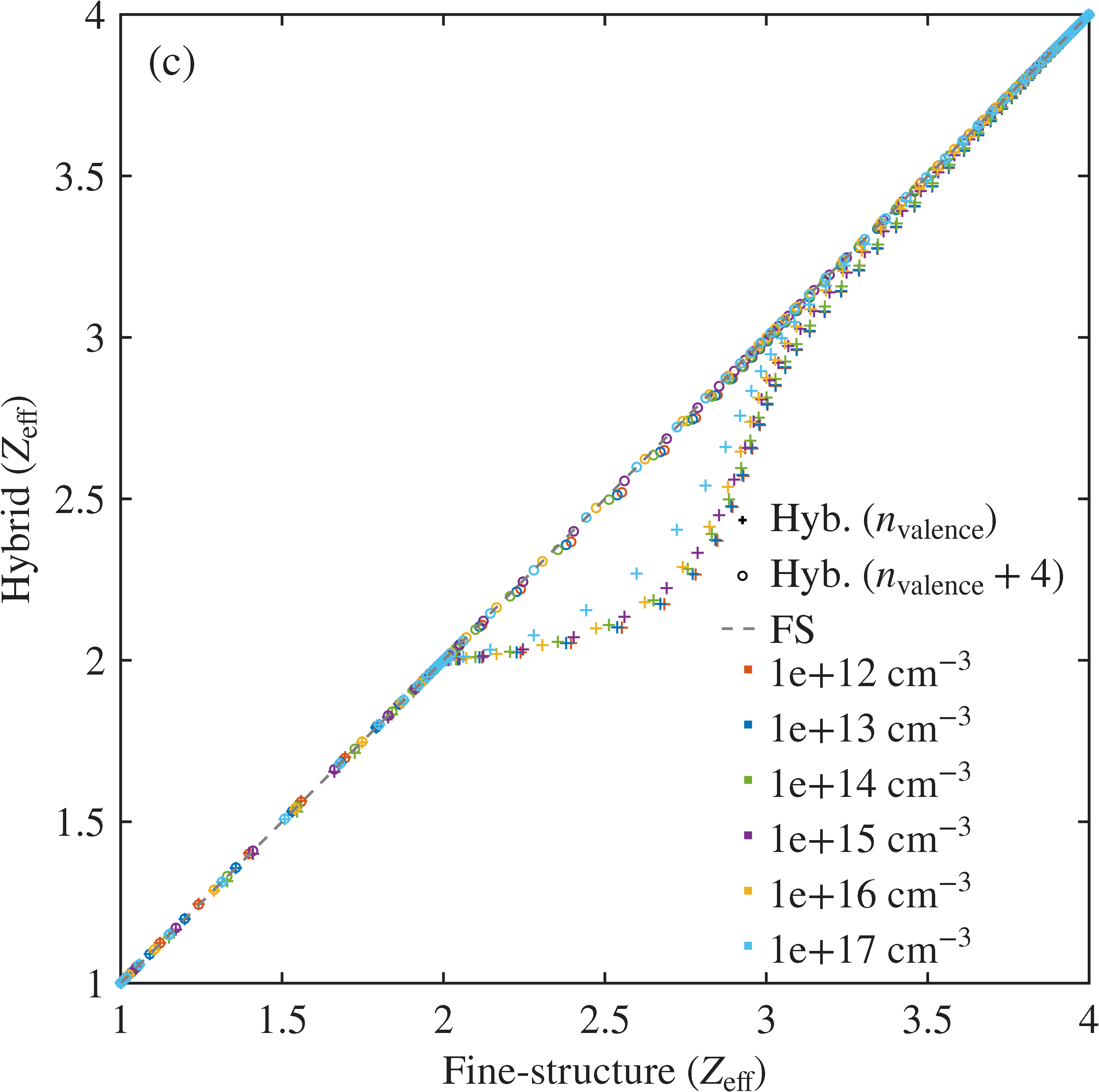}
\caption{\label{Hyb_Fine_Parity_Be}
\textbf{Beryllium:} Parity plots comparing (a) radiative power loss, $P_\mathrm{rad}$, (b) average charge state, $Z_\mathrm{avg}$, and (c) effective charge state, $Z_\mathrm{eff}$, predicted by Hybrid-($n_\text{valence}$) and Hybrid-($n_\text{valence}+4$) models against the fine-structure CR model across electron densities from $10^{12}$ to $10^{17}$~cm$^{-3}$. The dashed diagonal represents perfect agreement ($y = x$); points closer to the line indicate better agreement with the fine-structure benchmark.}

\end{figure}

The accuracy trends are further confirmed by the parity plots in 
Fig.~\ref{Hyb_Fine_Parity_Be}. The extended hybrid model shows a tight 
clustering along the diagonal across all temperatures and densities, indicating 
close agreement with the fully fine-structure benchmark. The reduced hybrid 
model exhibits a noticeably broader scatter, reflecting the loss of important 
fine-structure selectivity when averaging begins too close to the ground 
configuration.

Beyond these accuracy considerations, an important advantage of the
hybrid CR framework lies in its computational efficiency. In the case
of beryllium, the reduction from 9,790 fine-structure levels to 413
levels in the extended hybrid model (Hybrid-($n_{\text{valence}}+4$)),
and further down to 85 levels in the reduced version
(Hybrid-($n_{\text{valence}}$)), directly lowers the computational
effort required to solve the CR rate equations. Since the linear
algebra step scales approximately as $\mathcal{O}(N^{3})$, reducing
the size of the system produces a substantial decrease in solver time.

Furthermore, an equally important gain arises in the construction of
the rate matrix. All rate coefficients are generated from the
underlying cross-sections of the relevant atomic processes by
integrating over a Maxwellian electron distribution. The computational
cost of this step grows rapidly with the number of available states,
as the total number of transitions between states increases rapidly
with the size of the set of states. Reducing the number of active
states from nearly ten thousand to a few hundred thereby produces a
corresponding decrease in the number of transitions that must be
evaluated and assembled. This reduction leads to a significant
decrease in the computational cost of rate-matrix generation across
all plasma densities considered in this study.

A direct computational benchmark illustrates the magnitude of this improvement. For beryllium, the extended hybrid model provides an average reduction in the  total computation time, including both rate-matrix construction and the linear-system solution, by approximately a factor of twenty-five across all the electron densities considered in this work. Comparable reductions are observed for the lithium and helium cases, with average computational improvements of roughly eighteen-fold and six-fold, respectively. These results demonstrate that the hybrid framework provides consistently large efficiency gains across multiple plasma species.

For heavier ions, the computational advantage of the hybrid framework is expected to be even more pronounced. Detailed fine-structure CR modeling for such species may involve hundreds of thousands of levels and an extremely large number of level-to-level transitions, making direct calculations increasingly impractical. The scalability demonstrated for helium, lithium, and beryllium therefore  indicates that the hybrid approach is well suited for extending CR modeling to higher-$Z$ impurities, where maintaining fine-structure fidelity, while managing computational cost, is essential for accurate plasma diagnostics and integrated fusion modeling.

\section{Conclusion}
In this work we have developed and benchmarked hybrid CR models that combine fine-structure resolution with superconfiguration averaging, offering a practical balance between accuracy and computational efficiency. We considered two variants, Hybrid-($n_\text{valence}$) and Hybrid-($n_\text{valence}+4$). The first model retains detailed fine-structure levels up to $n_\text{valence}$ and the second up to $n_\text{valence}+4$, whereas higher-lying states (up to $n=10$) are statistically averaged into superconfiguration states. Comparisons with fully fine-structure CR models for helium, lithium, and beryllium show that Hybrid-($n_\text{valence}$) reproduces the overall trend of plasma observables, but introduces significant errors, particularly in radiation-dominated regimes, whereas Hybrid-($n_\text{valence}+4$) shows much closer agreement with fine-structure benchmarks, with fewer errors and a significant reduction in computational cost compared to the full fine-structure model. These results establish the hybrid CR framework as a versatile and scalable strategy for plasma modeling, allowing users to balance computational cost with the desired level of atomic-structure fidelity. The approach is well suited for fusion, astrophysical, and laboratory plasmas, where predictive accuracy must be maintained without additional computational overhead. Future studies will aim to extend and assess the hybrid approach for higher-$Z$ ions and for non-Maxwellian electron distributions, thereby broadening its applicability to more complex plasma environments.

\section{Acknowledgements}
This work was jointly supported by the US Department of Energy through
the Fusion Theory Program of the Office of Fusion Energy Sciences and
the SciDAC partnership on Tokamak Disruption Simulation between the
Office of Fusion Energy Sciences and the Office of Advanced Scientific
Computing, at Los Alamos National Laboratory (LANL) under Contract
No. 89233218CNA000001.  C.J.F, M.C.Z, and J.C would like to
specifically acknowledge Los Alamos National Laboratory (LANL) ASC PEM
Atomic Physics Project.  LANL is operated by Triad National Security,
LLC, for the National Nuclear Security Administration of the US
Department of Energy (Contract No. 89233218CNA000001).  This research
used resources of the National Energy Research Scientific Computing
Center, a DOE Office of Science User Facility supported by the Office
of Science of the U.S. Department of Energy under Contract
No. DE-AC02-05CH11231 using NERSC award FES-ERCAP0032298.

\bibliographystyle{apsrev}
\bibliography{reference.bib}

\end{document}